\documentclass[a4paper,twocolumn,11pt,unpublished]{quantumarticle}
\pdfoutput=1
\usepackage[utf8]{inputenc}
\usepackage[english]{babel}
\usepackage[T1]{fontenc}
\usepackage{amsmath}
\usepackage{hyperref}

\usepackage{tikz}
\usepackage{lipsum}

\usepackage{amssymb}
\usepackage{comment}
\usepackage{algorithm}
\usepackage{algorithmic}
\usepackage{xcolor}
\usepackage{braket}
\usepackage[numbers,sort&compress]{natbib}

\usepackage{qcircuit}
\usepackage{adjustbox}

\RequirePackage{xspace}
\newcommand{\tket}{\ensuremath{\text{t}\hspace{-0.5mm}\ket{\text{ket}}}\xspace}
\newcommand{\myopt}{{\sc Aqcel}\xspace}
\newcommand{\shigh}{\ensuremath{s_\epsilon^\text{high}}\xspace}
\newcommand{\smed}{\ensuremath{s_\epsilon^\text{med}}\xspace}
\newcommand{\slow}{\ensuremath{s_\epsilon^\text{low}}\xspace}
\newcommand{\sfrac}{\ensuremath{s_\epsilon^f}\xspace}
\newcommand{\porig}{\ensuremath{p^{\text{orig}}}}
\newcommand{\popt}{\ensuremath{p^{\text{opt}}}}
\newcommand{\Fsim}{\ensuremath{F_{\text{sim}}}}
\newcommand{\Fmeas}{\ensuremath{F_{\text{meas}}}}

\begin{document}

\title{Quantum Gate Pattern Recognition and Circuit Optimization for Scientific Applications}

\author{Wonho Jang}
\affiliation{Department of Physics, The University of Tokyo, 7-3-1 Hongo, Bunkyo-ku, Tokyo 113-0033, Japan}
\author{Koji Terashi}
\email{koji.terashi@cern.ch}
\orcid{0000-0001-6520-8070}
\affiliation{International Center for Elementary Particle Physics (ICEPP), The University of Tokyo, 7-3-1 Hongo, Bunkyo-ku, Tokyo 113-0033, Japan}
\author{Masahiko Saito}
\affiliation{International Center for Elementary Particle Physics (ICEPP), The University of Tokyo, 7-3-1 Hongo, Bunkyo-ku, Tokyo 113-0033, Japan}
\author{Christian W. Bauer}
\affiliation{Physics Division, Lawrence Berkeley National Laboratory, Berkeley, CA 94720, USA}
\author{Benjamin Nachman}
\affiliation{Physics Division, Lawrence Berkeley National Laboratory, Berkeley, CA 94720, USA}
\author{Yutaro Iiyama}
\affiliation{International Center for Elementary Particle Physics (ICEPP), The University of Tokyo, 7-3-1 Hongo, Bunkyo-ku, Tokyo 113-0033, Japan}
\author{Tomoe Kishimoto}
\affiliation{International Center for Elementary Particle Physics (ICEPP), The University of Tokyo, 7-3-1 Hongo, Bunkyo-ku, Tokyo 113-0033, Japan}
\author{Ryunosuke Okubo}
\affiliation{Department of Physics, The University of Tokyo, 7-3-1 Hongo, Bunkyo-ku, Tokyo 113-0033, Japan}
\author{Ryu Sawada}
\affiliation{International Center for Elementary Particle Physics (ICEPP), The University of Tokyo, 7-3-1 Hongo, Bunkyo-ku, Tokyo 113-0033, Japan}
\author{Junichi Tanaka}
\affiliation{International Center for Elementary Particle Physics (ICEPP), The University of Tokyo, 7-3-1 Hongo, Bunkyo-ku, Tokyo 113-0033, Japan}

\maketitle

\begin{abstract}
There is no unique way to encode a quantum algorithm into a quantum circuit.  With limited qubit counts, connectivities, and coherence times, circuit optimization is essential to make the best use of near-term quantum devices.  We introduce two separate ideas for circuit optimization and combine them in a multi-tiered quantum circuit optimization protocol called {\sc Aqcel}.  The first ingredient is a technique to recognize repeated patterns of quantum gates, opening up the possibility of future hardware co-optimization.  The second ingredient is an approach to reduce circuit complexity by identifying zero- or low-amplitude computational basis states and redundant gates.  As a demonstration, {\sc Aqcel} is deployed on an iterative and efficient quantum algorithm designed to model final state radiation in high energy physics.   For this algorithm, our optimization scheme brings a significant reduction in the gate count without losing any accuracy compared to the original circuit. Additionally, 
we have investigated whether this can be demonstrated on a quantum computer using polynomial resources.   
Our technique is generic and can be useful for a wide variety of quantum algorithms.
\end{abstract}

\section{Introduction}
\label{sec:intro}
Recent technology advances have resulted in a variety of universal quantum computers that are being used to implement quantum algorithms.  However, these noisy-intermediate-scale quantum (NISQ) devices~\cite{Preskill_2018} may not have sufficient qubit counts or qubit connectivity and may not have the capability to stay coherent for entirety of the operations in a particular algorithm implementation.  Despite these challenges, a variety of applications have emerged across science and industry.  For example, there are many promising studies in experimental and theoretical high energy physics (HEP) for exploiting quantum computers.  These studies include event classification~\cite{Mott:2017xdb,Zlokapa_2020,Chan:2019zwk,terashi2020event,Guan:2020bdl,belis2021higgs}, reconstructions of 
charged particle trajectories~\cite{Zlokapa:2019tkn,Tuysuz:2020ocw,Shapoval:2019txi,Bapst:2019llh} and physics objects~\cite{Wei:2019rqy,Das:2019hrw}, unfolding measured distributions~\cite{Cormier:2019kcq} as well as simulation of multi-particle emission processes~\cite{Bauer:2019qx,Nachman_2021}.  A common feature of all of these algorithms is that only simplified versions can be run on existing hardware due to the limitations mentioned above.

There are generically two strategies for improving the performance of NISQ computers to execute existing quantum algorithms. One strategy is to mitigate errors through active or passive modifications to the quantum state preparation and measurement protocols.   For example, readout errors can be mitigated through post-processing steps~\cite{Bauer:2019uf,bialczak_quantum_2010,neeley_generation_2010,dewes_characterization_2012,magesan_machine_2015,debnath_demonstration_2016,song_10-qubit_2017,gong_genuine_2019,wei_verifying_2020,havlicek_supervised_2019,chen_detector_2019,Chen_2020,maciejewski_mitigation_2020,urbanek_quantum_2020,nachman_unfolding_2020,hamilton_error-mitigated_2019,karalekas_quantum-classical_2020,geller_efficient_2020,geller_rigorous_2020,2010.07496} and gate errors can be mitigated by systematically enlarging errors before extrapolating to zero error~\cite{Dumitrescu:2018,PhysRevX.8.031027,PhysRevLett.119.180509,Kandala:2019,PhysRevA.102.012426,Otten_2019}.  A complementary strategy to error mitigation, that is the focus of this paper, is circuit optimization, also known as circuit compilation.  In particular, there is no unique way to encode a quantum algorithm into a set of gates, and certain realizations of an algorithm may be better-suited for a given quantum device.  One widely used tool is t|ket$\rangle$~\cite{Sivarajah_2020}, which contains a variety of architecture-agnostic and architecture-specific routines.  For example, Clifford identities such as $\text{CNOT}^2=\text{Identity}$ are automatically recognized.   There are also a variety of other toolkits for circuit optimization, including hardware-specific packages for quantum circuits~\cite{H_ner_2018,Green_2013,JavadiAbhari_2015,Svore_2018,Killoran_2019,gadi_aleksandrowicz_2019_2562111,smith2016practical,Steiger_2018,quantum_ai_team_and_collaborators_2020_4062499,mccaskey2018language,murali2019fullstack,robert_s_smith_2020_3677537,Nam_2018,venturelli2019quantum,murali2019noiseadaptive,Murali_2020,Peterson_2020,Leung_2017,Gokhale_2019,liu2020relaxed}.  Since t|ket$\rangle$ is a generic framework that contains many algorithms that have already been benchmarked against other procedures, it will serve as our baseline.

We introduce two techniques that can be used to optimize circuits and that are complementary to existing methods.  The first focuses on the identification of recurring sets of quantum gates in a circuit. Identifying such recurring sets of gates (RSG) can be very important, since any optimization of these RSGs has an enhanced effect on the overall circuit.  Furthermore, identifying recurring gate sets can be useful for future hardware optimizations where the fidelity of certain common operations can be enhanced at the expense of other, less frequent operations.  Depending on the operation(s), this optimization could be at the level of microwave pulse controls or it may require custom hardware architectures. 

The second technique optimizes a generic circuit by eliminating unnecessary gates or unused qubits such that the circuit depth becomes as short as possible.  One example where such an optimization can lead to simplifications is a case where a quantum circuit has been designed with complete generality in mind. In this case, for a certain initial state the circuit only reaches a selected set of intermediate states such that some operations become trivial and can be eliminated. 
The elimination of unnecessary gate operations introduced here focuses on controlled operations such as a Toffoli or a CNOT gate in a quantum circuit. 
The heart of the elimination technique resides in the identification of zero- or low-amplitude computational basis states, that
allows us to determine whether the entire gate or (part of) qubit controls can be removed.
Ref.~\cite{liu2020relaxed} proposed a similar technique to remove control gates with a quantum state analysis that identifies $X$-, $Y$- or $Z$-basis states. 
In addition, Ref.~\cite{liu2020relaxed} accounts for the basis states on target qubits to further simplify the circuit.
Our approach focuses only on $Z$-basis states on control qubits, but features a unique capability to perform the state determination using polynomial resources with a quantum hardware.

These two techniques are combined in an optimization protocol, called \textsc{Aqcel} (and pronounced ``excel'') for \textit{Advancing Quantum Circuit by \textsc{icEpp} and \textsc{Lbnl}}, and are 
presented in this paper.  
To demonstrate the effectiveness of the \myopt protocol, we will use a quantum algorithm from HEP to perform a calculation in Quantum Field Theory.  
The particular algorithm that we study models a \textit{parton shower}, which is the collinear final state radiation from energetic charged (under any force) particles~\cite{Nachman_2021}.  This algorithm is a useful benchmark because it provides an exponential speedup over the most efficient known classical algorithm and the circuit depth can be tuned for precision.  While we show results for this specific circuit, the proposed protocol has a wide range of applicability for quantum computing applications across science and industry.

This paper is organized as follows.  Section~\ref{sec:algo} provides an overview of the \myopt protocol.  The application of this protocol to the HEP example is presented in Sec.~\ref{sec:application}.  Following a brief discussion in Sec.~\ref{sec:discussion}, the paper concludes in Sec.~\ref{sec:conclusion}.

\section{\myopt optimization protocol}
\label{sec:algo}
\begin{figure}
\centering
\includegraphics[width=0.5\textwidth]{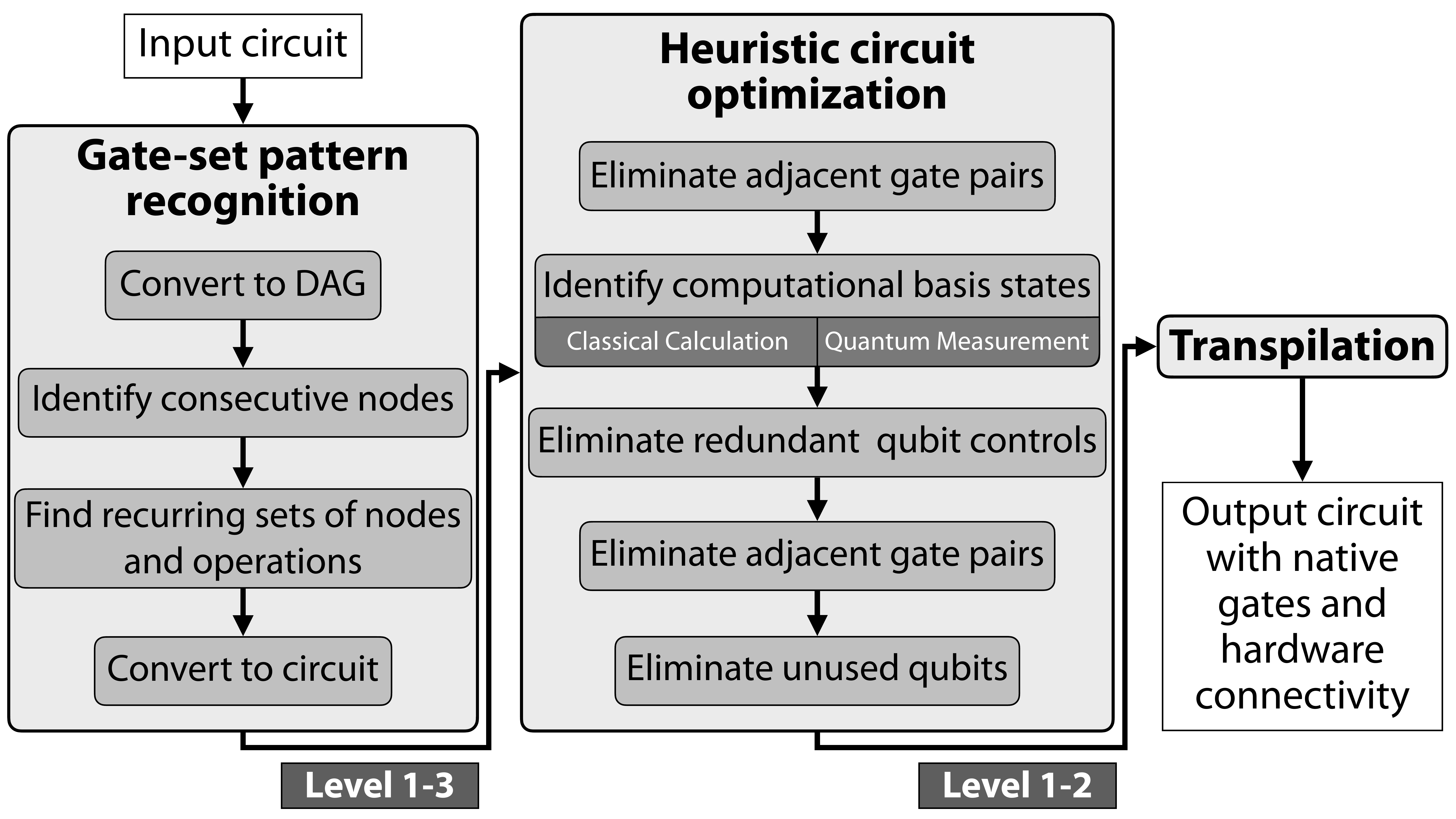}
\caption{Flowchart of the proposed optimization protocol.  The first part is the RSG pattern recognition, in which the circuit is converted into a directed acyclic graph~(DAG) to identify recurring quantum gates.  In the second part, we eliminate unnecessary gates and unused qubits through a heuristic approach.  Finally, the resulting circuit can be encoded into particular gates for specific hardware.}
\label{fig:flowchart} 
\end{figure}

As already mentioned, the \myopt protocol comprises two components: identification of recurring quantum gates (Sec.~\ref{subsec:gpr}) and elimination of unnecessary gates and unused qubits (Sec.~\ref{subsec:circ-opt}).  This approach focuses on circuit optimization at the algorithmic level, instead of at the level of a specific implementation using native gates for a particular quantum device.  A high-level flowchart for our protocol is presented in Fig.~\ref{fig:flowchart}.  The individual optimization steps are described below.

\subsection{Gate set pattern recognition}
\label{subsec:gpr}
First, the \myopt attempts to identify gate set patterns in an arbitrary quantum circuit and extract 
RSGs from the circuit. 

\begin{figure}
\centering
\includegraphics[width=0.45\textwidth]{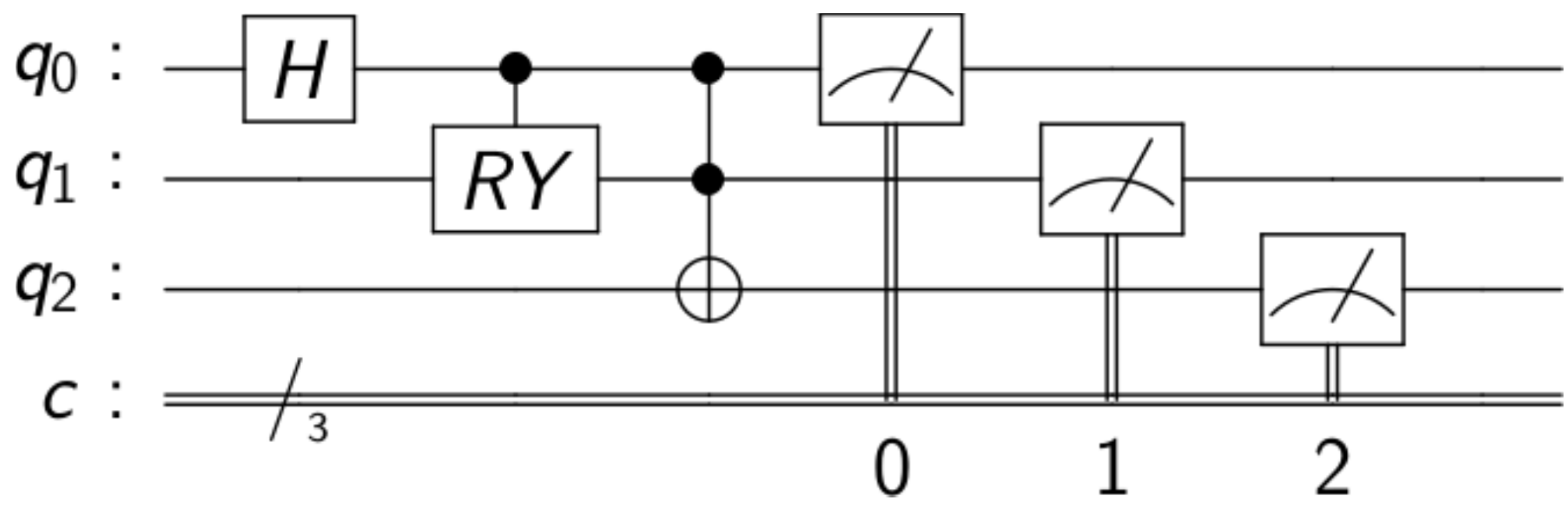}\\
\includegraphics[width=0.45\textwidth]{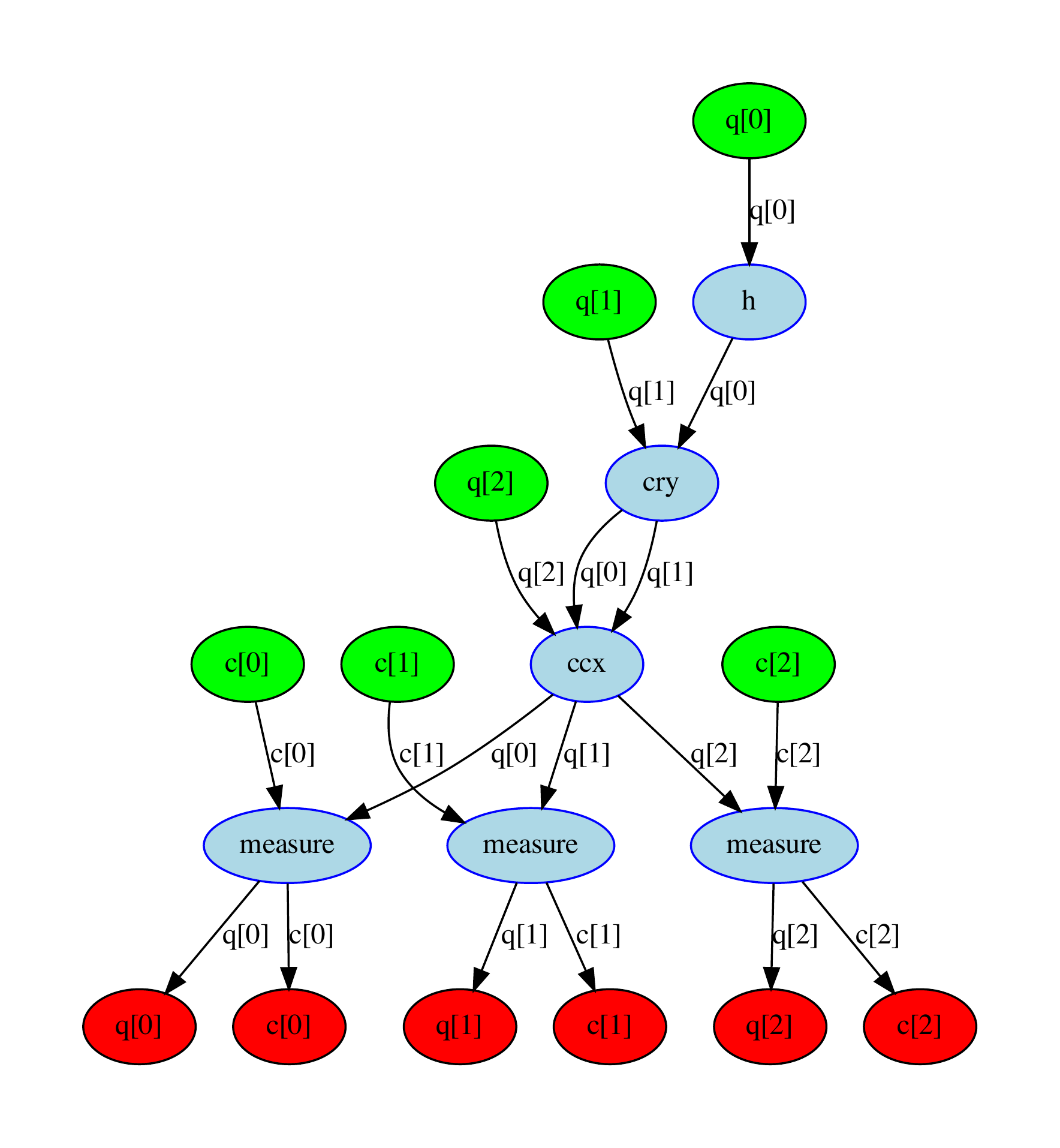}
\caption{An example circuit containing a Toffoli gate (top) and its corresponding DAG (bottom).}
\label{fig:toffoli_dag} 
\end{figure}

\subsubsection{Representation in directed acyclic graph}
\label{subsubsec:dag}
In a quantum circuit, individual qubits are manipulated sequentially by gate operations, meaning that 
the quantum state represented at a certain point of the circuit should not be affected by 
gate operations applied afterward (at a later point in the circuit). 
Such a structure can be described by a directed acyclic graph~(DAG).
A DAG allows us to easily check dependencies 
between qubits and extract a subset of the circuit that functions for certain tasks.

First, we convert a quantum circuit to the form of a DAG using the DAGCircuit class in Qiskit Terra API, 
where a node represents an operation by a quantum gate and an edge that connects the nodes represents a qubit.
In the case of a Toffoli gate, the node corresponding to the Toffoli gate has three incoming edges~(qubits before the gate operation) and three outgoing edges~(qubits after the gate operation).
Figure~\ref{fig:toffoli_dag} shows an example circuit containing a Toffoli gate and its corresponding DAG.

The gate set pattern recognition can be resolved through the DAG representation.
The identity of the RSG functionality can be ensured by checking the identity of DAGs of two circuits, as a graph isomorphism problem.
The algorithm of gate set pattern recognition consists of two steps:
(1) finding RSG candidates with DAG representation using 
depth-first search with heuristic pruning,
and (2) checking the DAG isomorphism by graph hashing with Weisfeiler Lehman graph hash~\cite{WLGraph}, as implemented in the NetworkX library~\cite{SciPyProceedings_11}.
The details of the gate set pattern recognition including computational complexity are given in Appendix~\ref{app:pattern_recognition_alg}, with the pseudocode of the algorithm.

\subsubsection{Tiered extraction of recurring gate sets}
\label{subsubsec:tiered-recog}
\begin{figure}
\centering
\includegraphics[width=0.5\textwidth]{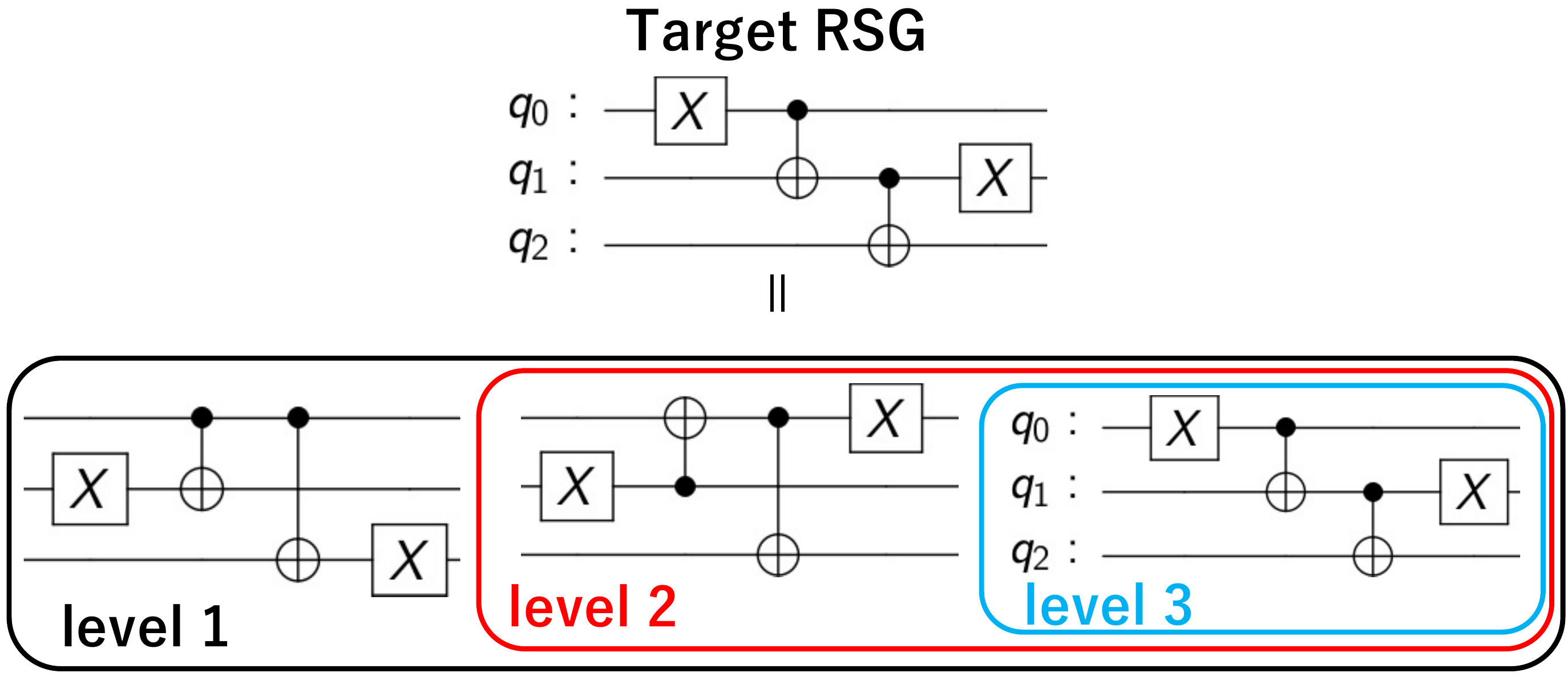}
\caption{Possible RSG patterns for a given target RSG corresponding to the three levels of matching criteria.}
\label{fig:rsg_level} 
\end{figure}

The appearance pattern of RSGs in a quantum circuit may depend on 
specific encoding of the quantum algorithm. To account for different patterns, 
we consider three different levels of matching criteria to define the recurrence of quantum gates:
\begin{description}
\item[{\bf Level 1} :] Only matching in gate types,\vspace{-2mm}
\item[{\bf Level 2} :] Matching in gate types and the {\it roles} of qubits that the gates act on,\vspace{-2mm}
\item[{\bf Level 3} :] Matching in gate types and both {\it roles} and {\it indices} of qubits that the gates act on.
\end{description}
The matching criterion in Level 1 is the least stringent: it just identifies the same sets of quantum gates appearing in the circuit,
irrespective of which qubits they act on. The Level 2 is more strict and ensures that the qubits the RSGs act on 
have the same roles. In other words, the qubit connections between the gates inside a single RSG are maintained 
but the qubit indices might vary between the RSGs. 
The Level 3 applies the most stringent condition, where the qubits that the RSGs act on 
must have the same roles and qubit indices, that is, the RSGs must appear on the identical set of qubits in the circuit. 
The appearance patterns of the RSGs are illustrated in Fig.~\ref{fig:rsg_level} 
for the three matching criteria.

The identified RSGs are ranked in terms of the product of the number of gates constituting the set and the number of occurrence of the set in the circuit. A specified number of top-ranked RSGs 
are extracted from the circuit in this step.

\subsection{Heuristic circuit optimization}
\label{subsec:circ-opt}
After attempting to identify RSGs in the circuit, a heuristic optimization procedure
takes place to make the circuit depth as short as possible by eliminating redundant gates or unused qubits.
In this step, we consider two levels of optimization:
\begin{description}
\item[{\bf Level 1} :] Optimize the entire circuit including RSGs,\vspace{-2mm}
\item[{\bf Level 2} :] Optimize the entire circuit, but for the RSGs only adjacent gate pairs (see below) are removed.
\end{description}
The Level 1 optimization would provide a shorter, more efficient circuit. 
Compared to Level 1, Level 2 optimization likely results in a deeper circuit for most cases, 
while it provides more room for 
improvement in later compilation stages if the RSGs have specialized low-level implementations.

\subsubsection{Basic idea of redundant controlled operations removal}
\label{subsubsec:basic_idea}
A controlled operation such as a CNOT or a Toffoli gate performs a different operation depending on
the quantum state of the system at the point where the gate is applied. Let $m$ be the number of control qubits of this operation. Consider expanding the state
of the full system $\ket{\psi}$ into a superposition of computational basis states as
\begin{equation}\label{eq:state_decomposition}
  \ket{\psi} = \sum_{j,k} c_{j,k} \ket{j}_{\text{ctl}} \otimes \ket{k},
\end{equation}
where $\ket{\cdot}_{\text{ctl}}$ denotes the state of the control qubits, while the unlabeled ket
corresponds to the rest of the system. 
We write the states as integers with $0 < j < 2^m-1$ and $0 < k < 2^{n-m} - 1$. We assume
that the controlled operation for the gate is applied when all control qubits are in the $\ket{1}$ state, which corresponds to the state $\ket{j}_{\text{ctl}} = \ket{11 \dots 1} = \ket{2^{m}-1}_{\text{ctl}}$. This allows one to classify the state of the system into three general classes using the amplitudes $c_{j,k}$:
\begin{description}
\item[\textbf{Triggering} :] $c_{j,k} \neq 0$ if and only if $j = 2^{m}-1$. The controlled operation of the gate in question is applied for all computational bases in the superposition. \vspace{-2mm}
\item[\textbf{Non-triggering} :] $c_{2^{m}-1,k} = 0$ for all $k$. The controlled operation is never applied.\vspace{-2mm}
\item[\textbf{Undetermined} :] The state is neither triggering nor non-triggering.
\end{description}

A circuit containing triggering or non-triggering controlled gates can be simplified by removing all
controls (triggering case) or by eliminating the gates entirely (non-triggering case). While an
undetermined single-qubit controlled gate cannot be simplified under the current scheme, an
undetermined multi-qubit controlled gate can be by removing the controls on some of the qubits, if
the state of the system satisfies the condition described in Appendix~\ref{app:qubit_control}.

As an example of this concept, consider the following simple circuit:
\begin{figure}[h!]
\centering
\leavevmode
\large
\Qcircuit @C=1em @R=1em @!R{
&&&\lstick{\ket{0}}  &\gate{H}  &\qw      &\qw      & \qw & \ctrl{1} & \qw  \\
&&&\lstick{\ket{0}}  &\ctrl{1}  &\gate{X} &\ctrl{1} & \qw & \ctrl{1} & \qw  \\
&&&\lstick{\ket{0}}  &\targ     &  \qw    &\targ    & \qw & \targ    & \qw  \\
}
\end{figure} \\
If the second qubit is in the initial state $\ket{0}$, the first CNOT gate has no effect and can be removed from the circuit as the $\ket{0}$ is the non-triggering state of CNOT. The second qubit before the second CNOT gate is in the state $\ket{1}$, which is the triggering state.
Therefore, the qubit control can be removed from the second CNOT gate.
The first two qubits before the Toffoli gate are in the superposition of $\ket{01}$ and $\ket{11}$, which is an undetermined state for the Toffoli gate.
Since the Toffoli gate has a triggering bitstring $\{11\}$, and the second qubit is always in the $\ket{1}$ state, 
this second qubit control can be removed from the Toffoli gate, replacing it with a CNOT gate controlled only on the first qubit.

The heuristic circuit optimization therefore requires, for each controlled gate, the identification of possible states the control qubits can take, and the removal of unnecessary parts of the controlled operations. These two steps are discussed in detail in the following.

It is well known that an arbitrary multi-qubit controlled-$U$ gate with $m$ control qubits can be decomposed into 
$\mathcal{O}(m)$ Toffoli and controlled-$U$ gates~\cite{PhysRevA.52.3457}. Therefore, 
in the remainder of this paper, we assume that all controlled gates are reduced to Toffoli gates denoted as $C^2[X]$, and singly-controlled unitary operation denoted as $C[U]$. This implies that the only triggering bitstrings we need to consider are either $\{1\}$ or $\{11\}$.
For a $n$-qubit circuit composed of $N$ multi-qubit controlled-$U$ gates, each having at most $n$ control qubits,
this decomposition results in at most $\tilde N = nN$ controlled gates.

\subsubsection{Identification of computational basis states}
\label{subsubsec:comp-base}
In general, a circuit consisting of $n$ qubits creates a quantum state described by a superposition of all of the $2^n$ computational basis states. However, it is rather common that a specific circuit produces a quantum state where only a subset of the computational basis states has nonzero amplitudes. Moreover, the number of finite-amplitude basis states depends on the initial state. This is why the three classes of the states of the system arise.

The state classification at each controlled gate can be determined either through a classical
simulation or by measuring the control qubits repeatedly. In the case of a
classical simulation, one can either perform the full calculation of the amplitudes, or simply track
all the computational basis states whose amplitudes may be nonzero at each point of the circuit without the calculation of the amplitudes.
\myopt adopts the latter method in the interest of the lowering the computational resource requirement.
When instead the quantum measurements
are used, the circuit is truncated right before the controlled gate in question, and the control
qubits are measured repeatedly at the truncation point. Finiteness of the relevant amplitudes can be
inferred from the distribution of the obtained bitstrings, albeit within the statistical uncertainty
of the measurements.

A few notes should be taken on the computational costs of the two methods. Consider an
$n$-qubit circuit with $N$ controlled gates. As discussed before, reducing this to either $C^2[X]$ or $C[U]$ results in ${\cal O}(\tilde N)$ single or double controlled gates. A classical simulation of the state vector before a given controlled gate has an exponential scaling in the number of qubits and requires
$\mathcal{O}(2^n)$ computations. On the other hand, measuring the $m=1$ or 2 control qubits $M$ times, which results in $M$ bitstrings of length $m$, only
requires $\mathcal{O}(M)$ operations. Repeating this for all $\tilde N$ gates requires $\mathcal{O}(\tilde N2^n)$
for the classical simulation and $\mathcal{O}(\tilde N^2M)$ when using quantum measurements. In other words, the number of operations for the bitstring determination with quantum measurements scales only polynomially with the number of qubits via a dependency through $\tilde{N} = nN$.

More details on the estimates of the computational resource necessary for the identification of computational basis states,
as well as other optimization steps, are described in Appendix~\ref{app:comp_resource}.

\subsubsection{Elimination of redundant controlled operations}
\label{subsubsec:ctrl-gate}
Once the nonzero-amplitude computational basis states are identified at each controlled gate, we remove the gate or its
controls if possible. When using classical simulation, the entire circuit is analyzed first before
the control elimination step. When quantum measurements are instead used, circuit execution,
measurements, and circuit optimization are performed separately at each controlled gate.

The control elimination step for each controlled gate proceeds as follows. For a $C[U]$ gate, compute the probability of observing $\ket{1}$ of the control qubit. If that probability is 1, eliminate the control and only keep the single unitary gate $U$. If the probability is 0, remove the controlled gate from the circuit. In all other cases, keep the controlled gate. For a $C^2[X]$ (Toffoli) gate, compute the probabilities of the four possible states $\ket{00}$, $\ket{01}$, $\ket{10}$, and $\ket{11}$. If the probability of $\ket{11}$ is 1, remove the two controls and only keep the $X$ gate. If the probability of $\ket{11}$ is 0, remove the entire Toffoli gate. If neither of those two conditions are true (the undetermined class), it is still possible to eliminate one of the two controls. This is true if the probability of the state $\ket{01}$ ($\ket{10}$) is zero, in which case one can eliminate the first (second) control.
The following pseudocode is the full algorithm for redundant controlled operations removal.

\noindent\makebox[\linewidth]{\rule{\columnwidth}{0.8pt}}
\textbf{Algorithm 1}: Redundant controlled operations removal

\vspace{-2mm}
\noindent\makebox[\linewidth]{\rule{\columnwidth}{0.4pt}}
\vspace{-6mm}
\begin{algorithmic}
\FORALL{$C[U]$ or $C^2[X]$ gate $g$ in the circuit}
    \STATE{execute circuit up to, but not including, $g$}
    \IF{$g$ is a $C[U]$ gate}
        \STATE{measure the control qubit $q$ in the $Z$ basis multiple times}
        \IF{$\{1\}$ is observed in the measurement results}
            \IF{$\{0\}$ is not observed in the measurement results}
                \STATE{turn $g$ into a $U$ gate acting on the target qubit}
            \ENDIF
        \ELSE
            \STATE{eliminate $g$}
        \ENDIF
    \ELSE
        \STATE{measure the control qubits $q_1 q_2$ in the $Z$ basis multiple times}
        \IF{$\{11\}$ is observed in the measurement results}
            \IF{neither $\{00\}$, $\{01\}$ nor $\{10\}$ is observed in the measurement results}
                \STATE{turn $g$ into an $X$ gate acting on the target qubit}
            \ELSIF{$\{01\}$ is not observed in the measurement results}
                \STATE{eliminate the control on $q_1$}
            \ELSIF{$\{10\}$ is not observed in the measurement results}
                \STATE{eliminate the control on $q_2$}
            \ENDIF
        \ELSE
            \STATE{eliminate $g$}
        \ENDIF
    \ENDIF
\ENDFOR
\end{algorithmic}
\vspace{-2mm}
\noindent\makebox[\linewidth]{\rule{\columnwidth}{0.4pt}}

Note that for noisy quantum circuits the measurements of the states will not be exact, and one expects errors in the probabilities to observe certain bitstrings. This means that one has to impose thresholds when deciding whether we call the state triggering, non-triggering or undetermined. Once such a threshold has been decided, the number of measurements required has to be large enough for the statistical uncertainty to be smaller than this threshold. This will be discussed in more detail in Sec.~\ref{sec:application} when we give explicit examples.

The computational cost of determining whether we can eliminate controls or the entire controlled operation is easily determined. Given the measured bitstrings, which as discussed in the previous section can be determined with $\mathcal{O}(\tilde N^2 M)$ operations,
one can compute the probabilities for each possible bitstring, and therefore decide whether to simplify a controlled operation using $\mathcal{O}(\tilde N)$ operations.
Some more details about the resource scaling are given in Appendix~\ref{app:comp_resource}.

Note that superfluous controlled operations can also be found and eliminated using the ZX-calculus~\cite{Coecke_2011,Duncan_2020}. In fact, the ZX-calculus is complete in the formal logic sense of the word, such that one can always prove that an unnecessary gate can be removed using the ZX-calculus. However, in general this scheme requires exponential resources, and therefore has no scaling advantage with respect to simply computing the state vectors. Nevertheless, the ZX-calculus is still incredibly powerful and underlies many of the optimization techniques of quantum transpilers, such as the \tket compiler we compare to later.

\subsubsection{Elimination of adjacent gate pairs}
\label{subsubsec:ident-gate}
Note that if a unitary operator $A$ and its Hermitian conjugate $A^\dag$ act on the same set of qubits adjacently, resulting in an identity operation, the gates implementing these operators can be removed from the circuit. While this is an obvious simplification, the removal of gates through the optimization steps described above can result in a circuit with such canceling gate pairs. For this reason, this step of gate reduction is applied before and after eliminating redundant controlled operations. 

\subsubsection{Elimination of unused qubits}
\label{subsubsec:unused-qubit}
After taking the above steps, the circuit is examined for qubits where no gate is applied at all. If found, such qubits can be safely removed from the circuit.
Such a situation occurs e.g., when
a quantum circuit designed to work universally with different initial states is executed using a 
specific initial state. An example of such a circuit is the sequential algorithm we consider in the next section.

\section{Application to quantum algorithm}
\label{sec:application}
The circuit optimization protocol described in Sec.~\ref{sec:algo} has been deployed to a quantum algorithm designed for HEP~\cite{Nachman_2021}.
The heuristic optimization (Sec.~\ref{subsec:circ-opt}) is performed at Level 1 for the optimization on existing quantum hardware.
In our results, we present how many gates are removed in three steps of the heuristic optimization, namely:
\begin{itemize}
\item Step 1: Eliminate adjacent gate pairs
\item Step 2: Eliminate redundant controlled operations
\item Step3: Eliminate adjacent gate pairs again
\end{itemize}

\subsection{Quantum parton shower algorithm}
\label{subsec:qalgo}
Simulating quantum field theories is a flagship scientific application of quantum computing.  It has been shown that a generic scattering process can be efficiently simulated on a quantum computer with polynomial resources~\cite{Jordan:2011ne}.  However, such circuits require prohibitive resources in the context of near-term devices.  

A complementary approach is to simulate one component of the scattering process.  In particular,  Ref.~\cite{Nachman_2021} proposed an algorithm to simulate the collinear radiation from particles that carry a nonzero fundamental charge.  Such radiation approximately factorizes from the rest of the scattering amplitude and can therefore be treated independently.  This factorization is the basis for parton shower Monte Carlo generators in HEP.  The quantum parton shower (QPS) algorithm provides an exponential speedup over known algorithms when the charge is not the same for all particles that can radiate. 

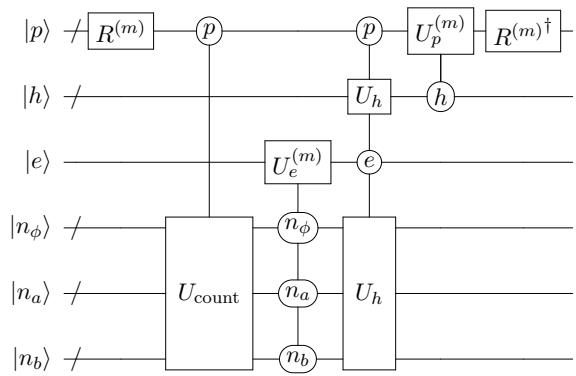
\begin{figure}
\centering
\resizebox{0.4\textwidth}{!}{
\Qcircuit @C=0.5em @R=0.8em @!R{
\lstick{\ket{p}} & {/} \qw &  \gate{R^{(m)}} &   \measure{\mbox{$p$}} \qwx[3]  & \qw  & \measure{\mbox{$p$}} \qwx[1]  &  \sgate{U^{(m)}_p}{1} &  \gate{{R^{(m)}}^{\dagger}}  & \qw   \\
\lstick{\ket{h}} & {/} \qw  & \qw & \qw & \qw & \sgate{U_{h}}{1}   &   \measure{\mbox{$h$}} \qwx[-1]  & \qw & \qw  \\ 
\lstick{\ket{e}} & \qw  & \qw & \qw & \gate{U^{(m)}_{e}} &  \measure{\mbox{$e$}} \qwx[1]   & \qw & \qw     & \qw   \\
\lstick{\ket{n_{\phi}}} & {/} \qw & \qw  & \multigate{2}{U_{\rm count}} & \measure{\mbox{$n_\phi$}} \qwx[-1]& \multigate{2}{U_{h}} & \qw  & \qw  & \qw     \\
\lstick{\ket{n_a}} & {/} \qw  & \qw &  \ghost{U_{\rm count}} & \measure{\mbox{$n_a$}} \qwx[-1]  &  \ghost{U_{h}} &  \qw & \qw  & \qw  \\ 
\lstick{\ket{n_b}} & {/} \qw  & \qw & \ghost{U_{\rm count}}  &\measure{\mbox{$n_b$}} \qwx[-1] & \ghost{U_{h}} & \qw & \qw   & \qw  
}
}
\caption{The $m$-th step of the quantum circuit for the algorithm proposed in Ref.~\cite{Nachman_2021}.  There are three physical registers: $\ket{p}$ containing the set of particles at this step; $\ket{h}$ for the branching history; and $\ket{e}$ which is a binary variable representing the presence or absence of an emission at this step.  The three lower registers count the number of particles of type $\phi$, $a$, and $b$ and are uncomputed before the end of the circuit. The exact form of the rotation matrices $R^{(m)}$ and the unitary operations $U_\text{count}$, $U_e^{(m)}$, $U_h$, and $U_p^{(m)}$ can be found in Ref.~\cite{Nachman_2021}.}
\label{fig:showercircuit}
\end{figure}

 The particular example demonstrated in Ref.~\cite{Nachman_2021} starts with $n$ fermions that can be either type $f_1$ or $f_2$.  These fermions can radiate a scalar particle $\phi$, which itself can split into a fermion-antifermion pair (of the same or different type).  The relevant parameters are the three couplings $g_1$, $g_2$, and $g_{12}$ between $f_1$ and $\phi$, $f_2$ and $\phi$, and $f_1\bar{f}_2$ ($\bar{f}_1f_2$) and $\phi$, respectively,
where antifermions are denoted by a bar above the fermion symbol $f$.
  The shower evolution is discretized into $N_\text{evol}$ steps and at each step, one of the particles could radiate/split or nothing happens.  This produces a precise result when $N_\text{evol}$ is large.  Figure~\ref{fig:showercircuit} shows the quantum circuit block for the $m$-th step of the quantum circuit.  First, the fermions are rotated into a new basis $f_a$ and $f_b$ where the effective mixing $g_{ab}$ between $f_a\bar{f}_b$ ($\bar{f}_af_b$) and $\phi$ is zero.  Then, the number of particles of each type is counted and stored in registers $n_a, n_b$, and $n_\phi$.  Next, a Sudakov factor is calculated to determine if an emission happens or not.  This operation depends only on the total number of particles of each type.  After the emission step, the particle and history registers are modified depending on the emission.  Lastly, the fermions are rotated back into the $f_1$ and $f_2$ basis.  Some of the steps in this algorithm are universal (independent of $m$) and some dependent on $m$ due to the running of coupling constants with the energy scale.

\subsection{Experimental setup}
\label{subsec:setup}
\begin{figure*}
\centering
\includegraphics[width=\textwidth]{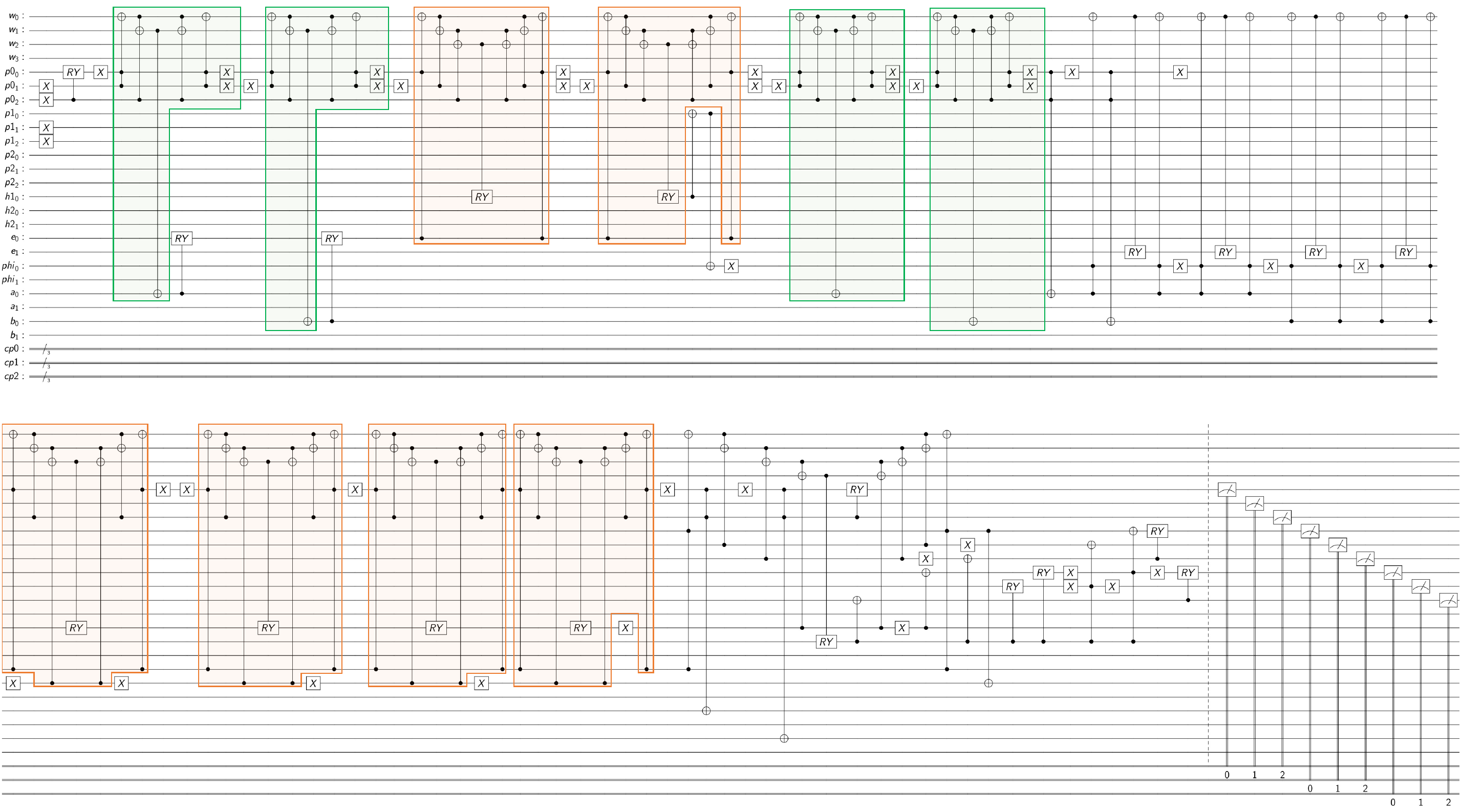}
\caption{Quantum circuit for the QPS simulation with two branching steps. 
The two RSGs, identified by the gate set pattern recognition step in the optimization scheme, 
are indicated by the ovals with different background colors.}
\label{fig:rsg_2step_ps} 
\end{figure*}

The QPS simulation is implemented into a quantum circuit using IBM Qiskit version~0.21.0~\cite{gadi_aleksandrowicz_2019_2562111}
with Terra~0.15.2, Aer~0.6.1 and Ignis~0.4.0 APIs in Python~3.8~\cite{10.5555/1593511}. 
First, we attempt to optimize the circuits running on a classical computer with a single 2.4~GHz Intel core i5 processor. 

In order to evaluate the \myopt performance, the same QPS circuit optimized using \tket in 
pytket~0.6.1 before transpilation is used 
as a reference. The optimization using \tket is done as follows. We consider the list of ten pre-defined passes~\footnote{The following 10 pre-defined passes are considered for the \tket optimization: {\it EulerAngleReduction(OpType.Rz,OpType.Rx)}, {\it RemoveRedundancies}, {\it GuidedPauliSimp}, {\it SquashHQS}, {\it FlattenRegisters}, {\it OptimisePhaseGadgets}, {\it KAKDecomposition}, {\it USquashIBM}, {\it CliffordSimp}, {\it FullPeepholeOptimise}.  Two more passes, {\it RebaseIBM}, {\it CommuteThroughMultis}, are also used once before selecting the pass from the list, which can be found at \texttt{https://cqcl.github.io/pytket/build/html/\\passes.html}.}.
The passes are tried one by one on the QPS circuit, and the one that reduces the number of gates the most is applied to the circuit. The same set of passes
are tried again on the resulting circuit to identify and apply the pass that most effectively reduces the gate count. 
This iterative process is repeated until the gate count is no longer reduced by any of the passes. The selected sequence of passes
is used for evaluating the \tket performance in the remainder of the studies.

The QPS algorithm is executed on the 27-qubit IBM's \textit{ibmq\_sydney} device, one of the IBM Quantum Falcon Processors, and the statevector simulator in Qiskit Aer with and without optimizing the circuit.
For the results obtained solely from the statevector simulator, all the qubits are assumed to be connected to each other 
(referred to as the ideal topology).
When executing the algorithm
on \textit{ibmq\_sydney}, the gates in the circuit are transformed into machine-native single- and two-qubit
gates, and the qubits are mapped to the hardware, accounting for the actual qubit connectivity.
For all the circuits tested with \textit{ibmq\_sydney} below, 
the noise-adaptive mapping is performed according to the read-out and CNOT 
gate errors from the calibration data as well as the qubit connection 
constraints~\footnote{This corresponds to the transpilation of level 3 pass manager, as implemented in Qiskit Terra.}.
Gate cancellations also take place at this stage using the commutativity of native gates and unitary synthesis, 
as documented in Qiskit Terra API. This qubit mapping and gate cancellation process are repeated eleven times, and
the circuit obtained with the smallest number of gates is finally tested with \textit{ibmq\_sydney}.

\subsection{Results}
\label{subsec:result}
\subsubsection{Circuit optimization for $N_\text{evol}=2$ branching steps using classical simulation}
\label{subsubsec:result_2step}

Circuit optimization performance of \myopt is evaluated for a quantum circuit of the QPS simulation
with $N_\text{evol}=2$ branching steps assuming an ideal topology.
The simulation does not consider any effects from hardware noise.
The initial state is chosen to be $\ket{f_1}$, and the coupling constants are set to 
$g_1=2$ and $g_2=g_{12}=1$. Both $f \to f'\phi$ and $\phi \to f\bar{f}$ processes are considered~\footnote{Ref.~\cite{Nachman_2021} noted that when these are unphysically removed, the circuit can be simulated efficiently classically (see also Ref.~\cite{Bauer:2019qx}).}.
The original circuit constructed using Qiskit is shown in Fig.~\ref{fig:rsg_2step_ps}.

First, the RSG pattern recognition is performed against the circuit.
When the Level 2 RSG pattern recognition is applied, two RSGs are identified, as also shown in Fig.~\ref{fig:rsg_2step_ps}, 
with the requirements on the number of nodes in each RSG being between 5 and 7
and the number of repetitions being 4 or more. If the matching level is raised from
Level 2 to 3, candidate patterns with smaller numbers of nodes or repetitions are generally found.

Next, the heuristic optimization (Sec.~\ref{subsec:circ-opt}) is performed over the entire circuit
at Level 1.  This step consists of identifying nonzero-amplitude computational basis
states, removing redundant controlled operations, removing adjacent canceling gate pairs (performed twice),
and removing unused qubits. Nonzero-amplitude computational basis states are identified through classical calculation.

\begin{figure}[ht]
\centering
\includegraphics[width=0.48\textwidth]{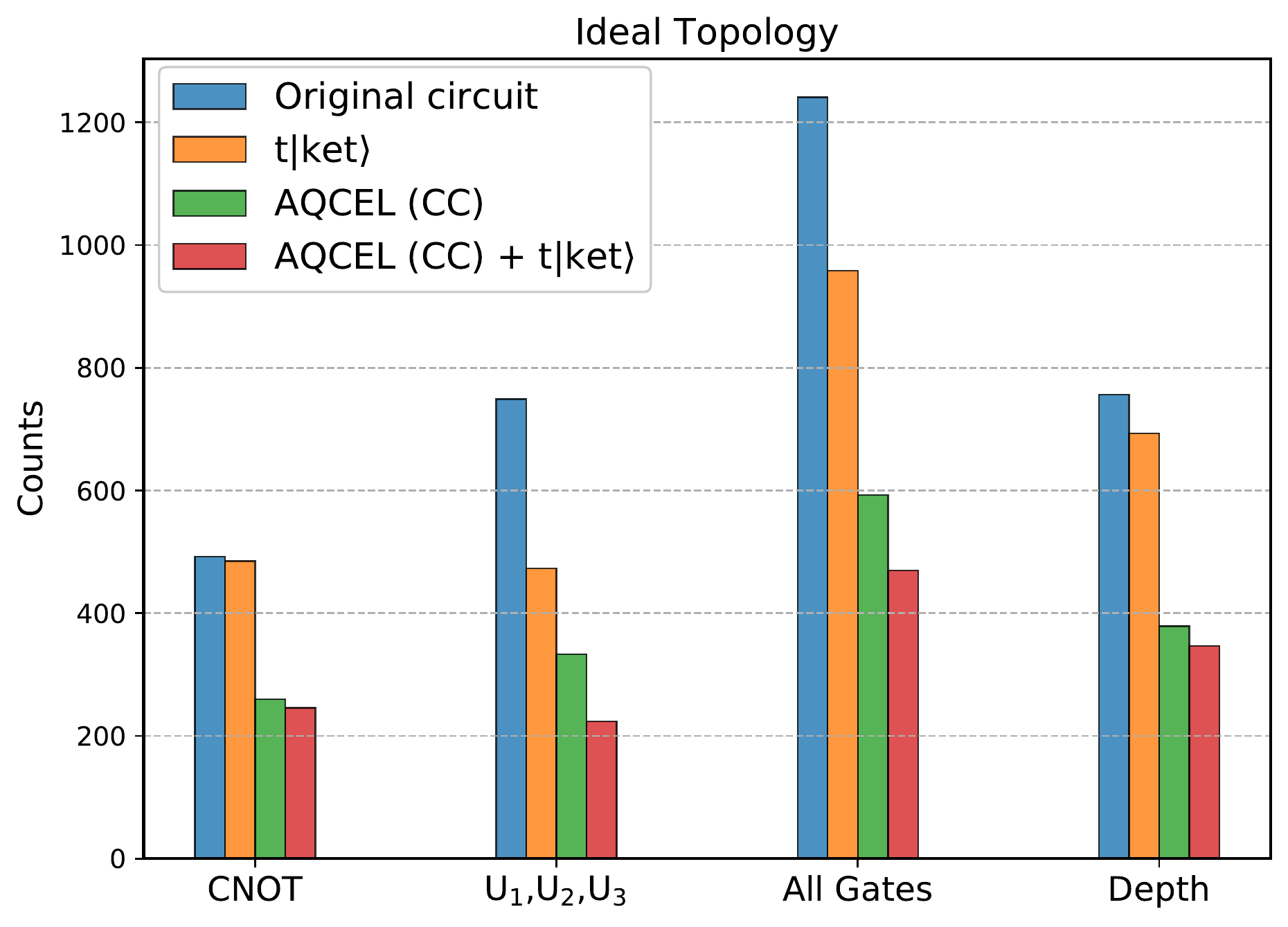}
\caption{Numbers of single-qubit ($U_{1,2,3}$) gates, CNOT gates and the sum of the two 
as well as the depth of the two-branching step QPS circuit decomposed into native gates before and after optimization. 
The computational basis states with nonzero amplitudes at controlled gates
are identified using classical calculation in the heuristic optimization step of \myopt.}
\label{fig:gatecounts_2step_ps} 
\end{figure}

After the algorithmic level circuit optimization, the quantum gates in the circuit 
are decomposed into single-qubit gates ($U_1$, $U_2$, $U_3$) and CNOT gates. 
Figure~\ref{fig:gatecounts_2step_ps} shows
the numbers of the single-qubit and CNOT gates, the sum of the two, and the depth of the circuit before and after the optimization.
The circuit depth is defined as the length of the longest path from the input to the measurement gates, with 
each gate counted as a unit, as implemented in Qiskit.
The figure compares the values from the original circuit, the circuit optimized with \tket only, that with \myopt only, and that with the combination of the two.
The \myopt optimizer reduces the total number of gates by 52\%, resulting in a 50\% reduction of the circuit depth. In particular, the reduction of the number of CNOT gates
is 47\%. This compares to \tket, which reduces the total number of gates by 23\%, CNOT by 1\%, and the circuit depth by 8\%. This means that, for the QPS algorithm,
\myopt is 38\% more efficient than \tket in reducing the gate counts, and 46\% more specifically for CNOT, and makes the circuit
45\% shorter.
Combination of the two optimizers is even more effective; a sequential application of \myopt and \tket reduces the gate count by 62\% (50\% for CNOT only) and the depth by
54\% with respect to the original circuit. In other words, the combined optimizer is 51\% more efficient than
the \tket alone for gate reduction (49\% for CNOT only), producing a 50\% shorter circuit.

For the \myopt optimizer, the gate reduction occurs mostly at the stage where the redundant qubit controls are removed. 
Starting with 1241 gates (excluding barrier and measurement gates), the first adjacent gate-pair elimination, the redundant qubit control reduction, and the second gate-pair elimination steps remove 132, 510 (41\% of the 1241 gates), and 6 gates, respectively.
In terms of the computational cost, the wall time is by far dominated by the two adjacent gate-pair elimination steps combined, accounting for 98\% of the total time,
followed by a sub-dominant contribution of 1\% from the redundant qubit control reduction. 

Finally, the number of qubits is reduced from 24 to 21 with the \myopt optimizer, while it is unchanged by \tket.
One qubit is removed from each of the three registers $n_a, n_b$, and $n_\phi$  
because those qubits are used only for $N_\text{evol}\geq3$ branching steps.

\subsubsection{Circuit optimization for $N_\text{evol}=1$ branching step using classical simulation}
\label{subssubsec:result_1step_sim}

The quantum circuit for the two-branching step QPS simulation is still too deep to produce useful results on
a real existing quantum computer, even after optimizing the circuit. Therefore, we consider the circuit with only
one branching step using the \textit{ibmq\_sydney} and the statevector simulator. 
The initial state, coupling constants, and considered processes are the same as those used for the 
$N_\text{evol}=2$ branching steps simulation.

First, we examine the gate and qubit counts for the one-branching step QPS simulation assuming an ideal topology.  Starting with 472 gates, the \myopt optimizer removes 10, 346
(73\% of 472 gates), and 2 gates in the three steps of the heuristic optimization, in the order
given above. The adjacent gate-pair elimination step still dominates the wall time (97\%). However,
the redundant qubit control reduction now takes about 3 times less time than that for the
two-branching step simulation, which is consistent with the exponential behavior of the computing cost of the step,
as discussed in Sec.~\ref{sec:algo}. The number of qubits is reduced from 15 to 13 with the \myopt optimizer.
One of four ancilla qubits is removed because three ancillas are sufficient for decomposing all the multi-controlled gates 
in the $N_\text{evol}=1$ step.
The register $n_\phi$, composed of only one qubit, is also removed because it is used only for the case where 
the initial state is $\ket{\phi}$.

\begin{figure}
\centering
\includegraphics[width=0.48\textwidth]{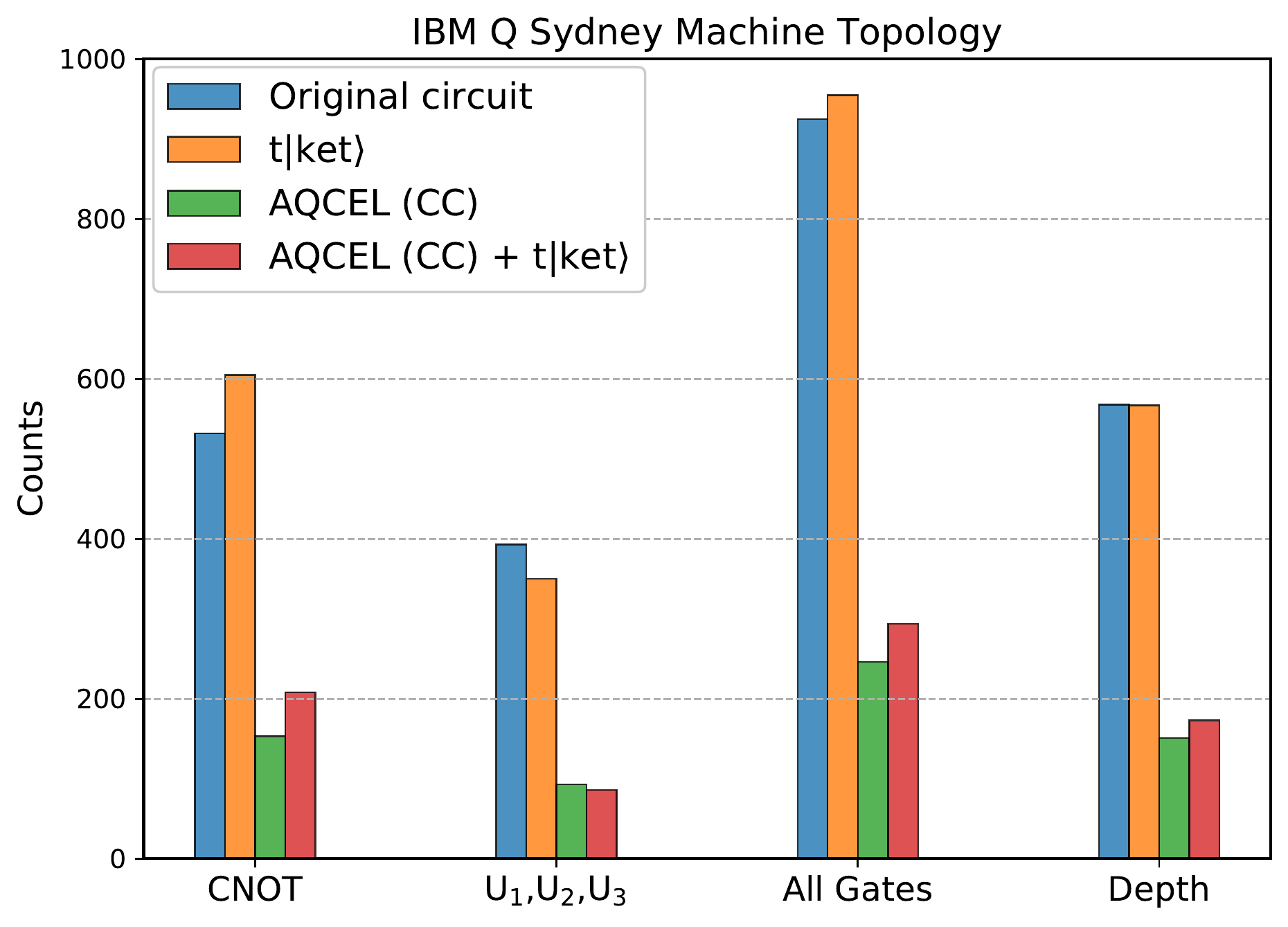}
\caption{Numbers of single-qubit ($U_{1,2,3}$) gates, CNOT gates and the sum of the two 
as well as the depth of the one-branching step QPS circuit transpiled considering \textit{ibmq\_sydney} topology before and after the optimizations.
The computational basis states with nonzero amplitudes at controlled gates are identified using classical calculation
in the heuristic optimization step of \myopt.}
\label{fig:gatecounts_1step_ps} 
\end{figure}

Next, the optimized circuits are transpiled considering the qubit connectivity of \textit{ibmq\_sydney}.
Figure~\ref{fig:gatecounts_1step_ps} shows the same set of distributions as in 
Fig.~\ref{fig:gatecounts_2step_ps}, but for the one-branching step QPS simulation with \textit{ibmq\_sydney}-specific transpilation.
The \myopt optimizer achieves a significant reduction of native gates for the one branching step as well.
The relative reduction is more drastic for the one branching step than the two branching steps,
mainly because the former (shallow) circuit has relatively more zero-amplitude computational basis states than the 
latter (deep) circuit.

\subsubsection{Circuit optimization for $N_\text{evol}=1$ branching step using quantum measurements}
\label{subssubsec:result_1step_measure}

Now we evaluate the performance of the optimizers using a quantum hardware. A particular challenge when
employing \myopt with a real quantum computer is in the determination of the bitstring probabilities
of the control qubits at each controlled gate using quantum measurements.  Due to hardware noise,
the list of observed bitstrings would contain contributions from errors on the preceding gates and
the measurement itself.

To mitigate the measurement errors, we obtain the correction by measuring 
the calibration matrix for the control qubits (with 8192 shots per measurement) using Qiskit Ignis API. 
The correction is then applied to the observed distribution with a least-squares fitting approach.

The errors incurred by gate imperfection accumulate throughout the circuit execution and degrade the performance. 
In particular, the CNOT gate error is the most significant source of the degradation.
To mitigate the effects from CNOT errors due to depolarizing noise, we employed a zero-noise extrapolation (ZNE) technique with identity insertions, first proposed in Ref.~\cite{Dumitrescu:2018} and generalized in Ref.~\cite{PhysRevA.102.012426}. 
The Fixed Identity Insertion Method of Ref.~\cite{PhysRevA.102.012426} amplifies the CNOT errors by replacing the 
$i$-th CNOT gate in a circuit with $2n_{i}+1$ CNOT gates and extrapolating the measurements to the limit of zero error. 
In the \myopt protocol with the QPS simulation circuit, each CNOT gate is replaced with 3 CNOT gates ($n_i=1$).

To account for remaining contributions to the measurements from gate errors, we opt to ignore the observed bitstrings with occurrence below certain thresholds (called cutoff thresholds). 
This is justified under the assumption that the residual gate errors act as a perturbation, inserting spurious computational basis states with small amplitudes into the superposition of the system.

In order to choose the cutoff thresholds, we consider errors in the single-qubit gates ($U_{1,2,3}$) and CNOT gates separately for all the hardware qubits.
The reported error rates at the time of the experiment, measured during the preceding calibration run of the hardware, are used for the calculations.
Let the $U_{1,2,3}$ and CNOT error rates be $\epsilon_{\mathrm{U}}^{(i)}$ and $\epsilon_{\mathrm{CX}}^{(i,j)}$, respectively, with $i$ and $j$ indicating qubits that the gates act on.
We can approximate the probabilities, $p_{\mathrm{U}}$ and $p_{\mathrm{CX}}$, of measuring the states 
without any $U_{1,2,3}$ or CNOT gate errors occurring anywhere in the circuit by performing qubit-wise (index-dependent) multiplications 
of the error rates:
\begin{align}
p_{\mathrm{U}} &= \prod_{i} \left(1-\epsilon_{\mathrm{U}}^{(i)}\right)^{n_{\mathrm{U}}^{(i)}},\\
p_{\mathrm{CX}} &= \prod_{i\neq j} \left(1-\epsilon_{\mathrm{CX}}^{(i,j)}\right)^{n_{\mathrm{CX}}^{(i,j)}},
\end{align}
where $n_{\mathrm{U}}^{(i)}$ and $n_{\mathrm{CX}}^{(i,j)}$ are the numbers of $U_{1,2,3}$ and CNOT gates acting on the corresponding qubits, respectively.
The probability $p_\epsilon$ of measuring the states with at least one gate error occurring anywhere in the circuit is
\begin{align}
p_\epsilon &= 1-p_{\mathrm{U}}p_{\mathrm{CX}}
\nonumber\\
&\sim N_{\mathrm{CX}} \epsilon_{\mathrm{CX}}.
\end{align}
In the last approximation, we have assumed that all CNOT errors are equal, much larger than single gate errors but still much smaller than one: $\epsilon_{\mathrm{U}}^{(i)} \ll \epsilon_{\mathrm{CX}}^{(i,j)} = \epsilon_{\mathrm{CX}} \ll 1$.
Applying the ZNE to mitigate the depolarizing CNOT errors, the $p_\epsilon$ is reduced to $p_\epsilon^{\rm{zne}}$:
\begin{align}
p_\epsilon^{\rm{zne}} &= 1 - \left(\frac{3}{2}p_{\mathrm{CX}} - \frac{1}{2}p_{\mathrm{CX}}^3 \right)
\nonumber\\
&\sim N_{\mathrm{CX}}^2 \epsilon_{\mathrm{CX}}^2
\end{align}
by ignoring the contributions from single-qubit gate errors.

The first cutoff threshold is chosen to be
\begin{equation}
\shigh := p_\epsilon^{\rm{zne}}.
\end{equation}
This corresponds to making an extreme assumption that any gate error during circuit
execution would result in a specific bitstring observed at the measurement, and attempting to
discard that bitstring. The second threshold:
\begin{equation}
\slow := p_\epsilon^{\rm{zne}} / 2^m,
\end{equation}
where $m$ is the number of the measured control qubits, corresponds to another extreme assumption
that the gate errors would result in a uniform distribution of all possible bitstrings. The third and final threshold is the average of the above two:
\begin{equation}
\smed := (\slow + \shigh) / 2.
\end{equation}

It should be noted that $p_\epsilon^{\rm{zne}}$ increases as the circuit execution proceeds, because $p_\epsilon^{\rm{zne}}$ accounts for
the ZNE-mitigated error rates of all the preceding gates in the circuit. 
As an alternative strategy to these dynamic cutoff thresholds, we also examine 
the static thresholds, \sfrac, that are kept constant throughout the circuit, with the values between 0.05 and
0.3.
We also consider capping the dynamic thresholds of \slow, \smed and \shigh at 0.2, with the reason explained later.

\begin{figure}
\centering
\includegraphics[width=0.48\textwidth]{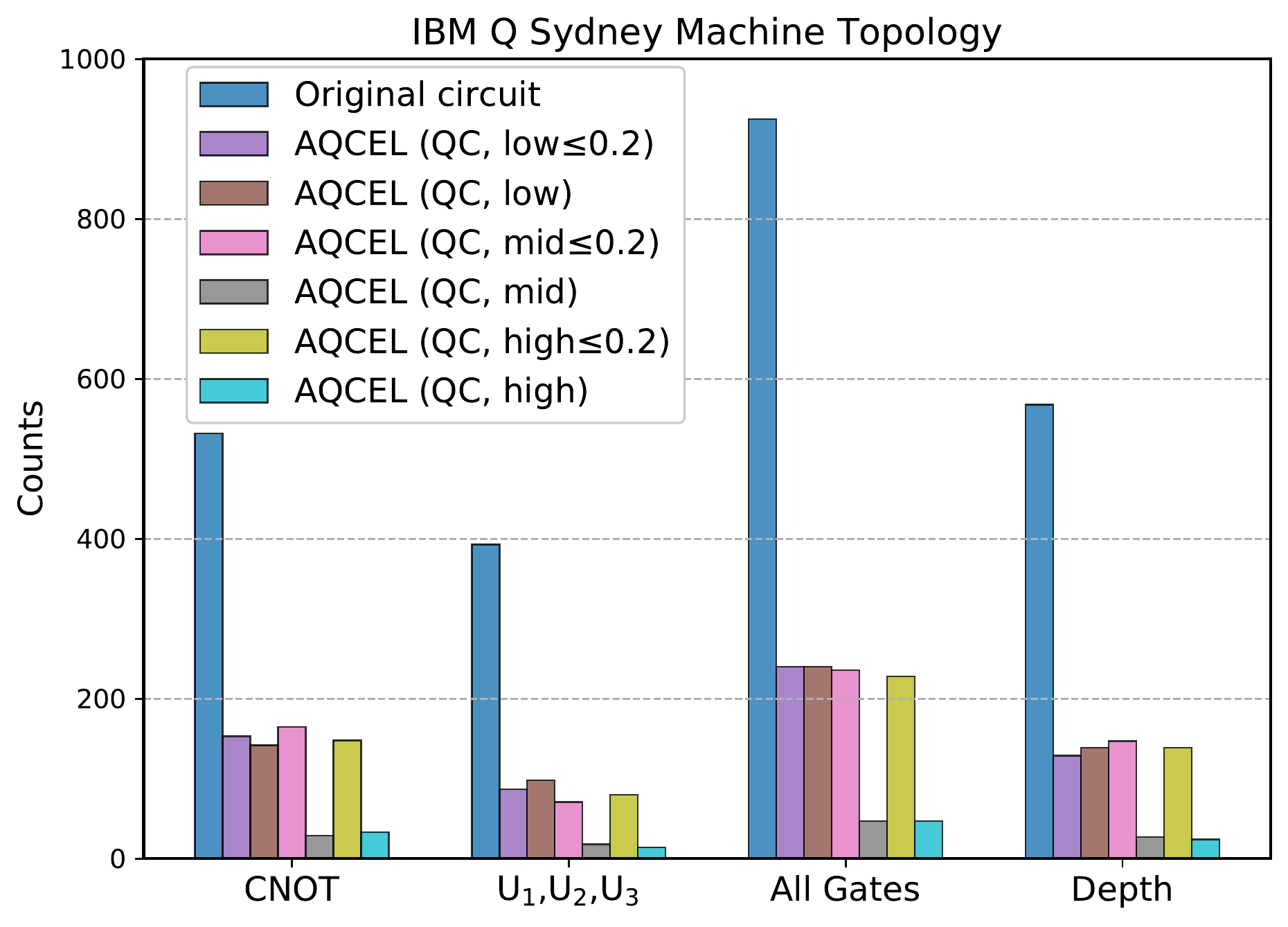}
\caption{Numbers of single-qubit ($U_{1,2,3}$) gates, CNOT gates and the sum of the two 
as well as the depth of the one-branching step QPS circuit transpiled considering \textit{ibmq\_sydney} topology before and after optimization. 
The probabilities of observing various bitstrings in the control qubits are measured using \textit{ibmq\_sydney}
in the heuristic optimization step, and the three dynamic cutoff thresholds of \slow, \smed and \shigh are applied.}
\label{fig:gatecounts_1step_ps_qc1} 
\end{figure}

Discarding all bitstrings with occurrence under certain thresholds obviously introduces errors of
its own. For example, we observe that discarding bitstrings using the unbounded \shigh as the threshold
for the one-branching step QPS simulation circuit results in an elimination of most of the
controlled gates in the later part of the circuit, rendering the circuit practically meaningless. Therefore,
the actual cutoff threshold of \myopt should be selected by considering the trade-off between the
efficiency of the circuit optimization and the accuracy of the optimized circuit~\footnote{In the
actual implementation, the threshold of 0.05 is applied to all
the three cases to suppress contributions from imperfect measurement error mitigation.}.

\begin{figure}
\centering
\includegraphics[width=0.48\textwidth]{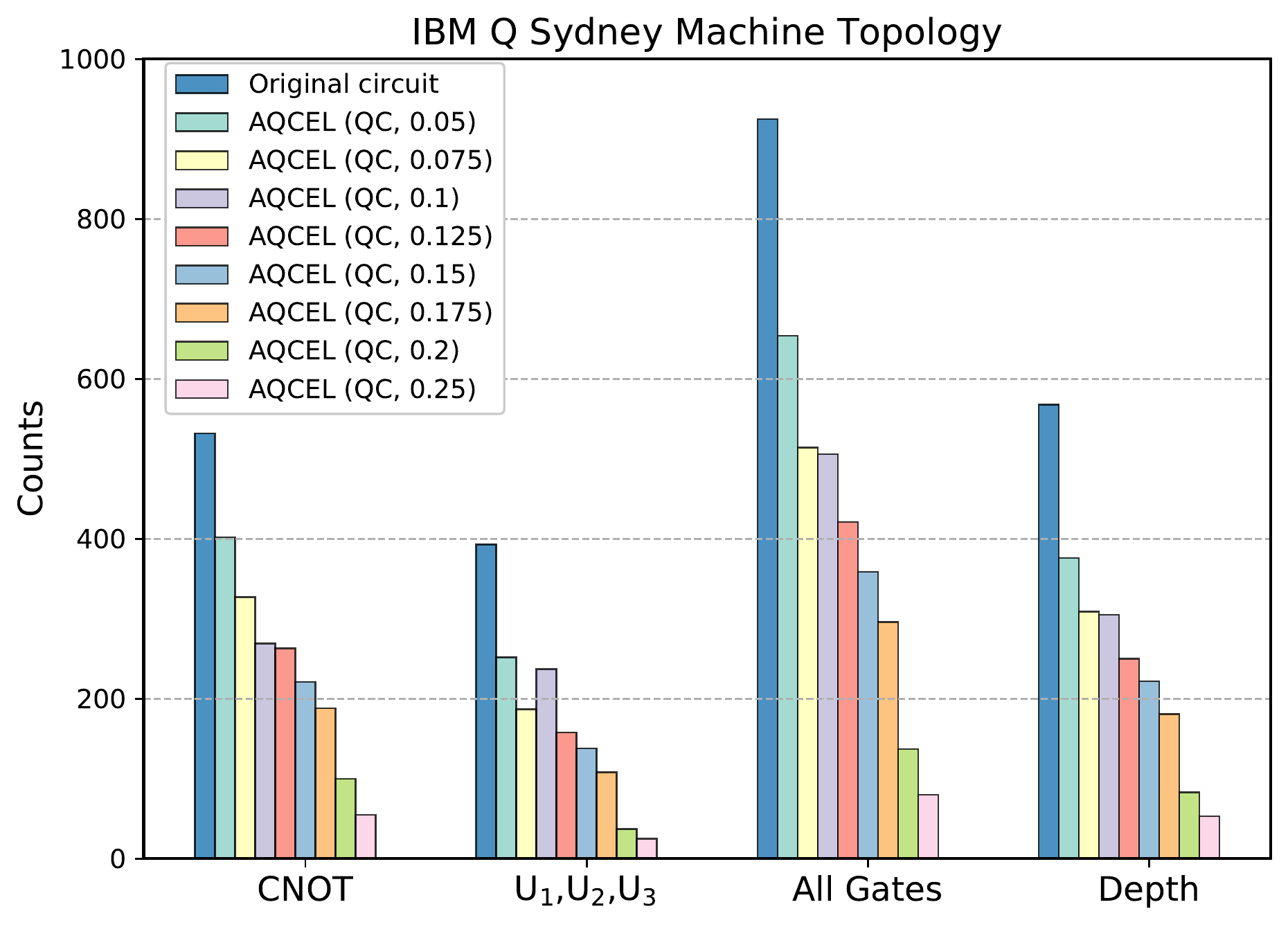}
\caption{Numbers of single-qubit ($U_{1,2,3}$) gates, CNOT gates and the sum of the two 
as well as the depth of the one-branching step QPS circuit transpiled considering \textit{ibmq\_sydney} topology before and after optimization.
The probabilities of observing various bitstrings in the control qubits are measured using \textit{ibmq\_sydney}
in the heuristic optimization step, and the static cutoff thresholds of \sfrac are applied.}
\label{fig:gatecounts_1step_ps_qc2} 
\end{figure}

Figure~\ref{fig:gatecounts_1step_ps_qc1} shows the gate counts obtained from \myopt optimizations
using actual measurements on \textit{ibmq\_sydney} under the dynamic cutoff thresholds.
The gate counts decrease as the threshold is raised from \slow to \shigh, as expected.
Figure~\ref{fig:gatecounts_1step_ps_qc2} shows the same distributions obtained with the static thresholds. Almost no gate survives under the threshold of 0.3, likely
implying a significant loss of accuracy for the computation result.

The number of qubits is reduced from 15 to 13 under all the dynamic thresholds. Under the static thresholds, the number of qubits is reduced from 15 to 13 for
$0.05 \leq \sfrac \leq 0.2$, but a significant reduction to 8 is seen for $\sfrac = 0.3$.

To evaluate the accuracy of the optimized circuit, we consider a classical fidelity of the final state of
the circuit, which is defined in terms of the probability distribution of the bitstrings
observed in the measurement at the end of the circuit. This quantity, denoted as $F$ and referred to
as just ``fidelity'' hereafter, is given by
\begin{equation}
F = \sum_k \sqrt{\porig_k \popt_k},
\end{equation}
where the index $k$ runs over the bitstrings. The quantities $\porig_k$ and $\popt_k$ are the
probabilities of observing $k$ in the original and optimized circuits, respectively.

In fact, we compute two fidelity values for each optimization method. The first, denoted $\Fsim$, aims to quantify the amount of modifications to the original circuit introduced by the optimization 
procedure at the algorithmic level. To calculate $\Fsim$, both {\porig} and {\popt} are computed using the statevector simulation. The value of $\Fsim = 1$ indicates that the optimized circuit is identical to the original circuit (up to a possible phase difference on each of the qubits), while a deviation from unity gives a measure of how much the optimization has modified the circuit.  

The second fidelity value, $\Fmeas$, is computed using measurements from actual quantum computer for $\popt$. The $\popt$ is estimated from the rate at which a bitstring occurs in a large number of repeated measurements. 
The {\porig} is computed using simulation as for the $\Fsim$.
Even if the optimized circuit is identical to the original circuit, the presence of noise will mean $\Fmeas < 1$, with the difference from unity getting larger when more gates (particularly CNOT gates) are present in the circuit. Removing CNOT gates to obtain the optimized circuit will lower the overall effect of noise and raise the $\Fmeas$ value. However, 
in some cases the CNOT gate removal would affect low-amplitude computational basis states, making the optimized circuit different from the original circuit, hence suppress 
the {\Fmeas} value. Thus, the {\Fmeas} is a measure that reflects the tradeoff of making the circuit shorter and changing the circuit through optimization.

\begin{figure}
\centering
\includegraphics[width=0.48\textwidth]{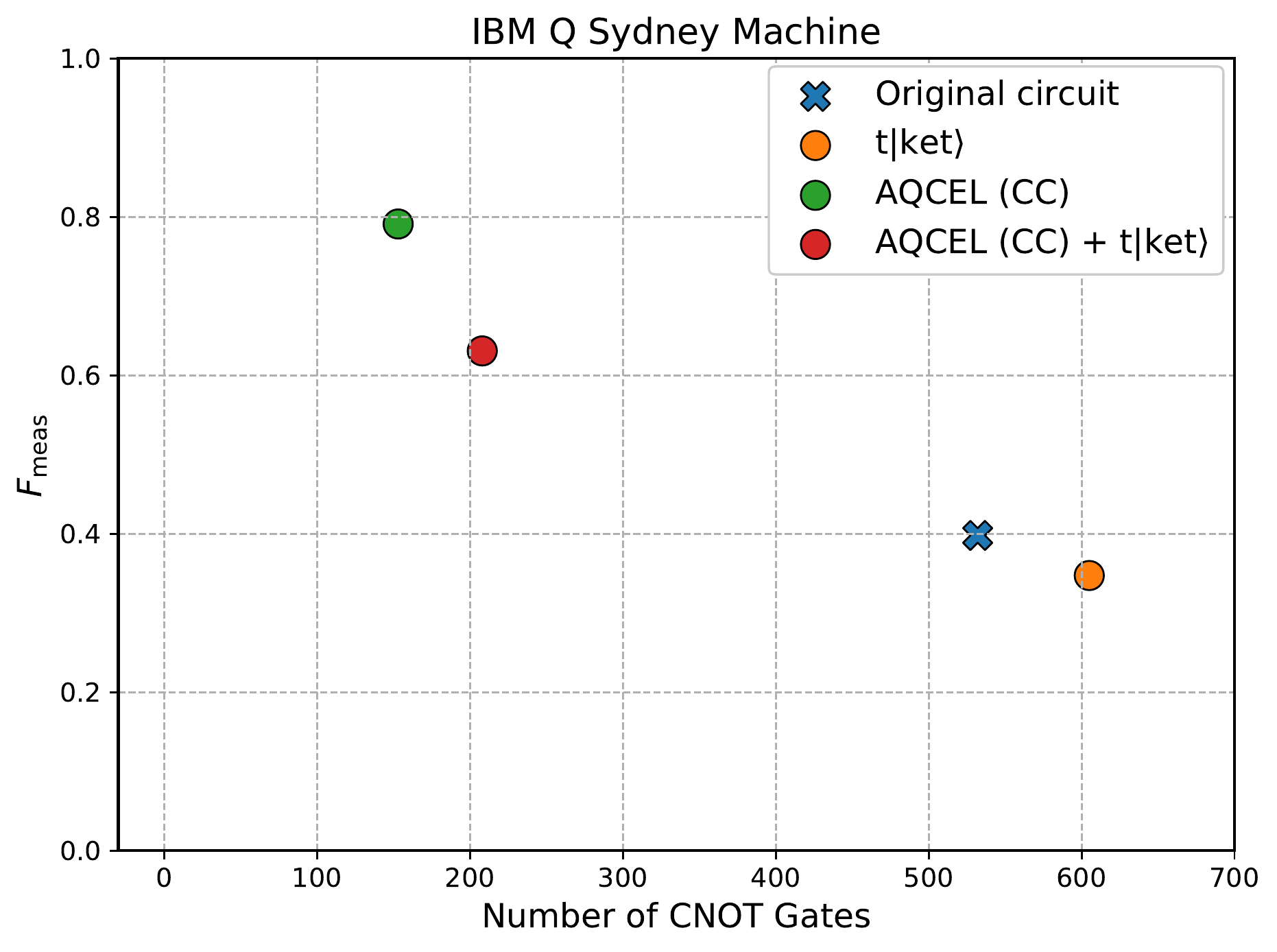}
\caption{Fidelity {\Fmeas} versus the number of CNOT gates for the one-branching step QPS circuit transpiled considering \textit{ibmq\_sydney} topology before and after optimization. The computational basis states with nonzero amplitudes at controlled gates are identified using classical calculation in the heuristic optimization step. These transpiled circuits are executed on \textit{ibmq\_sydney} to obtain the {\Fmeas}.
}
\label{fig:fidelity_1step_ps} 
\end{figure}

\begin{figure}[ht]
\centering
\includegraphics[width=0.48\textwidth]{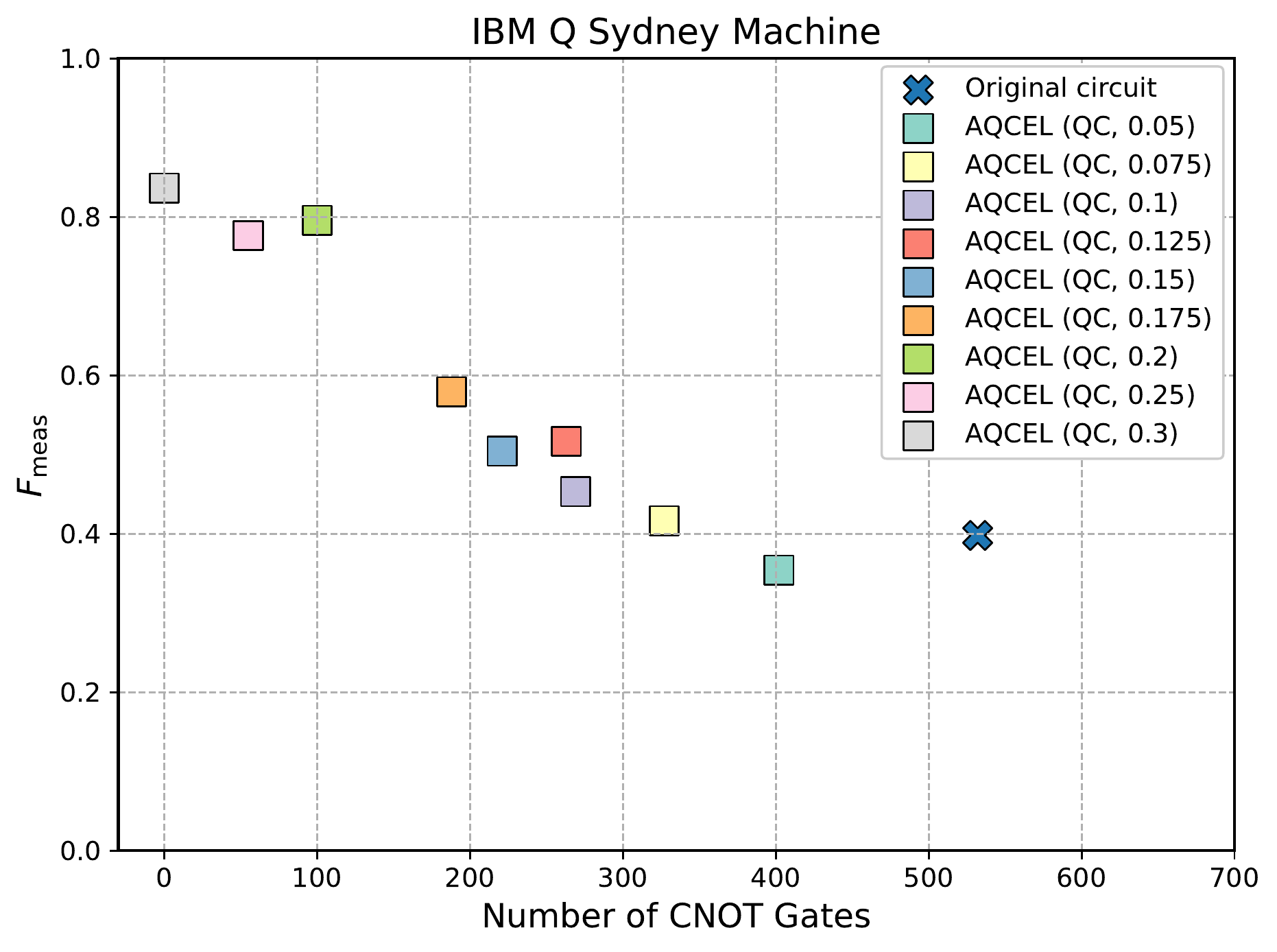}\\
\includegraphics[width=0.48\textwidth]{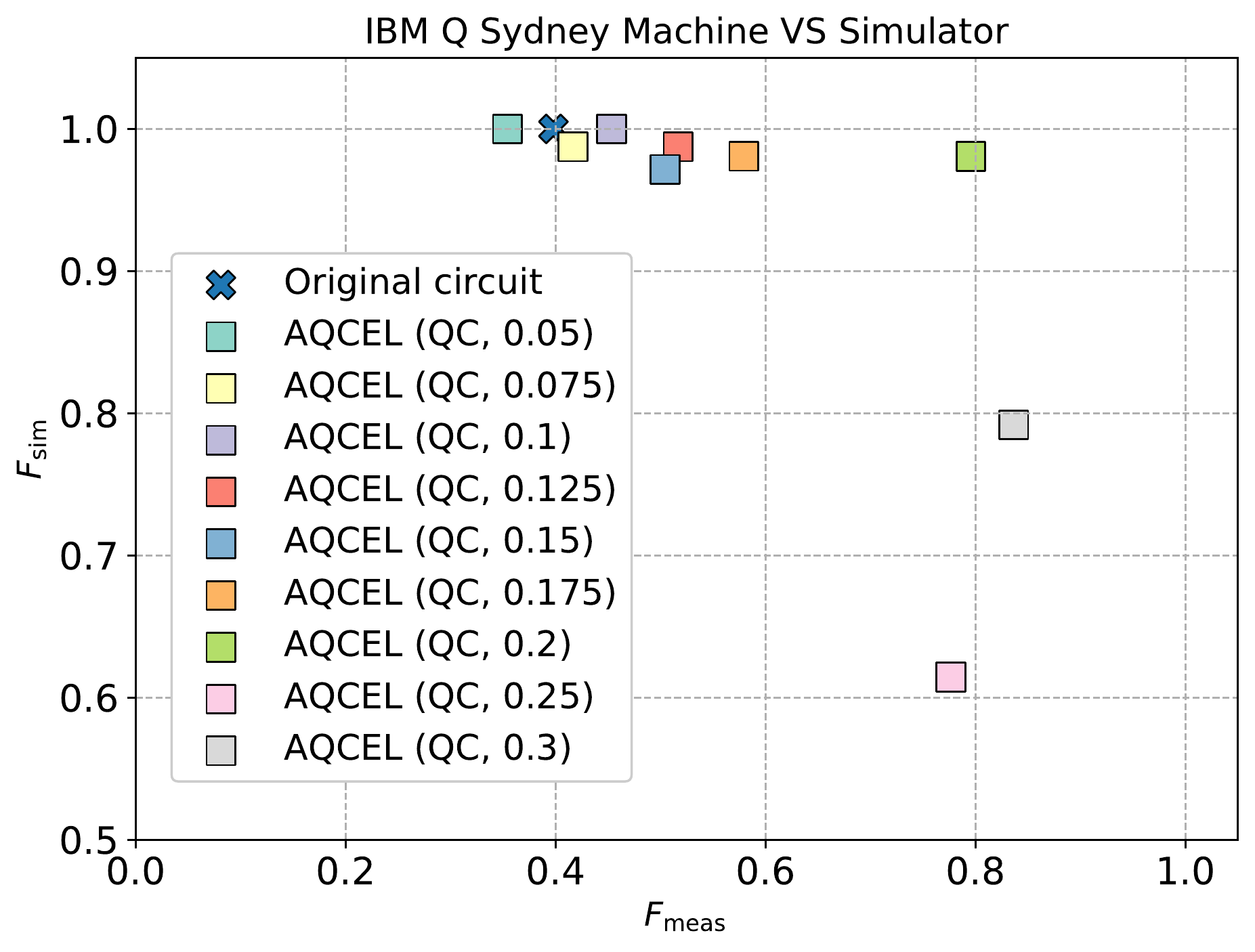}
\caption{Fidelity {\Fmeas} versus the number of CNOT gates (top) and fidelities {\Fmeas} versus {\Fsim} (bottom) for the one-branching step QPS circuit transpiled considering 
\textit{ibmq\_sydney} topology before and after optimization.
The probabilities of observing various bitstrings in the control qubits are measured using \textit{ibmq\_sydney}
in the heuristic optimization step, and the static thresholds of \sfrac are applied.
These transpiled circuits are executed on \textit{ibmq\_sydney} to obtain the {\Fmeas} and a statevector simulator to obtain the \Fsim.
}
\label{fig:fidelity_1step_ps_qc2} 
\end{figure}

Figure~\ref{fig:fidelity_1step_ps} shows the fidelity {\Fmeas} versus the number of CNOT gates 
before and after optimization, where the optimization is performed
using the classical simulation~\footnote{The {\Fmeas} value as a function of the number of all gates including $U_{1,2,3}$ shows the same trend as that in the {\Fmeas} versus the CNOT gate counts. This confirms that the {\Fmeas} value is predominantly determined by CNOT error contributions to the bitstring probabilities of $\popt$.}. One can see that shortening the circuit with less CNOT gates increases the {\Fmeas} as expected. 
The {\Fsim} values stay at unity for all the optimized circuits (not shown), validating that the optimization does not affect the computational 
accuracy with respect to the original circuit.
The measurements are performed 81,920 times for each of the circuit to obtain 
the {\Fmeas} values, and measurement error mitigation is not used in these and the following {\Fmeas} measurements.

When the elimination of redundant qubit controls is performed based on measurements using a quantum computer with the static thresholds \sfrac, the {\Fmeas} versus CNOT gate counts 
become those shown in Fig.~\ref{fig:fidelity_1step_ps_qc2}. Also shown in the figure is the correlation between {\Fsim} and {\Fmeas}.
We observe that the {\Fmeas} increases with increasing \sfrac value up to $\sfrac=0.3$. However, the {\Fsim} stays close to unity up to $\sfrac=0.2$ then 
decreases significantly, signaling that the optimized circuit becomes too far from the original circuit with $\sfrac>0.25$.
For the circuit considered here, the optimization performance therefore appears to be the best with $\sfrac \sim 0.2$.
The relations between {\Fmeas} and gate counts have been compared with and without applying the ZNE for the static thresholds. It shows that the {\Fmeas} improves with ZNE at 
low \sfrac thresholds below $\sim 0.15$, indicating that the accuracy of the optimized circuit improves by discarding spurious low-amplitude basis states 
with the suppression of CNOT errors.

\begin{figure*}[ht]
\centering
\includegraphics[width=0.48\textwidth]{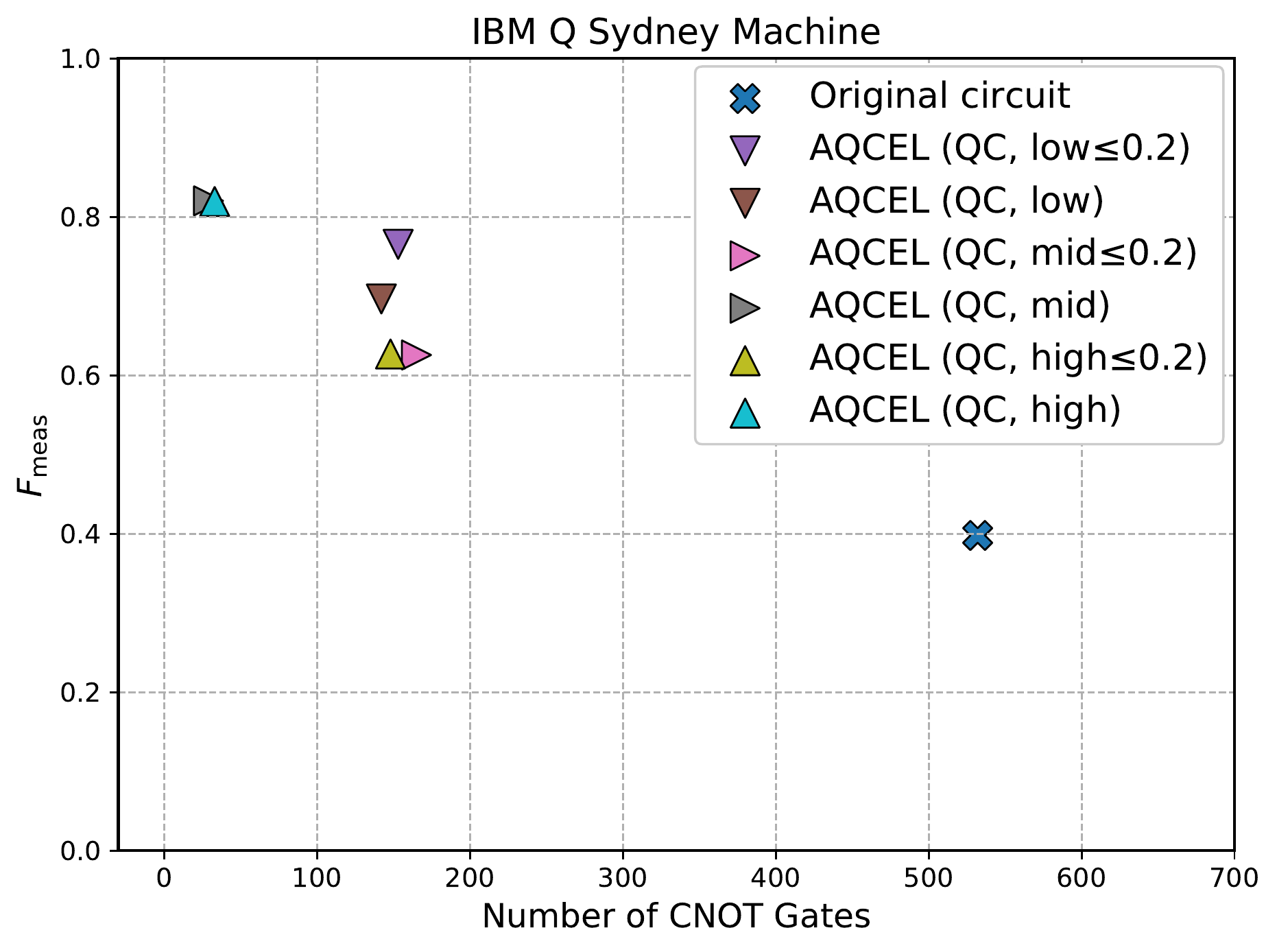}
\includegraphics[width=0.48\textwidth]{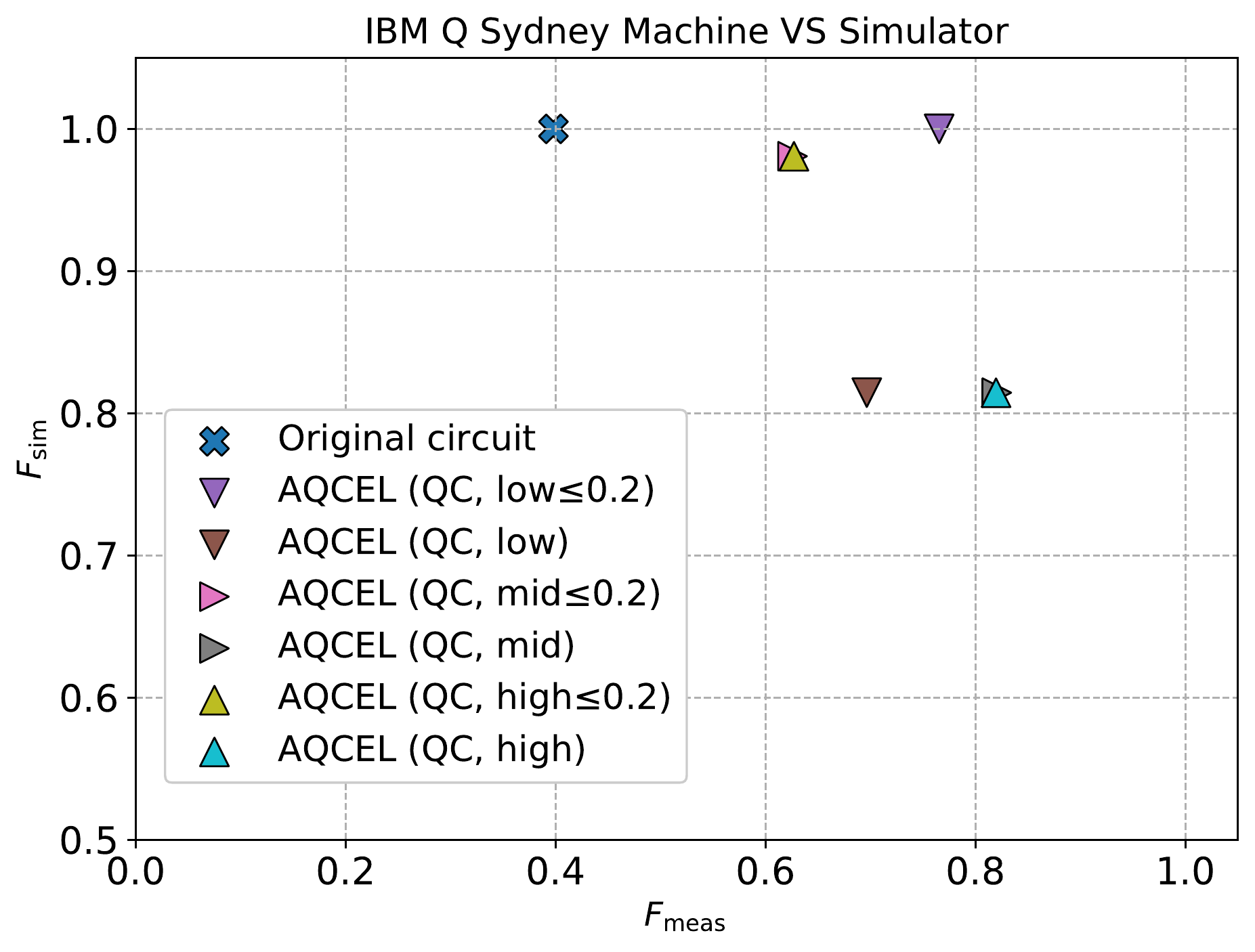}
\caption{Fidelity {\Fmeas} versus the number of CNOT gates (left) and fidelities {\Fmeas} versus {\Fsim} (right) for the one-branching step QPS circuit transpiled considering 
\textit{ibmq\_sydney} topology before and after optimization.
The probabilities of observing various bitstrings in the control qubits are measured using \textit{ibmq\_sydney}
in the heuristic optimization step, and the three dynamic thresholds of \slow, \smed and \shigh are applied.
These transpiled circuits are executed on \textit{ibmq\_sydney} to obtain the {\Fmeas} and a statevector simulator to obtain \Fsim.}
\label{fig:fidelity_1step_ps_qc1} 
\end{figure*}

In Fig.~\ref{fig:fidelity_1step_ps_qc1} we show the results of the optimization with the dynamic thresholds of \shigh, \smed and \slow. 
The results for the capped variants, where the threshold is capped at 0.2, are also shown. 
The {\Fmeas} generally improves with higher thresholds, but the {\Fsim} gets significantly worse for all the three thresholds without capping. 
The capped variants leave more gates in the circuit and have lower {\Fmeas} than the unbounded cases. However, they can restore the 
computational accuracy, making the {\Fsim} values much closer to unity. An exception is the case of \slow where the {\Fmeas} value is unchanged or slightly better
with capping.

\begin{figure}[htbp]
\centering
\includegraphics[width=0.48\textwidth]{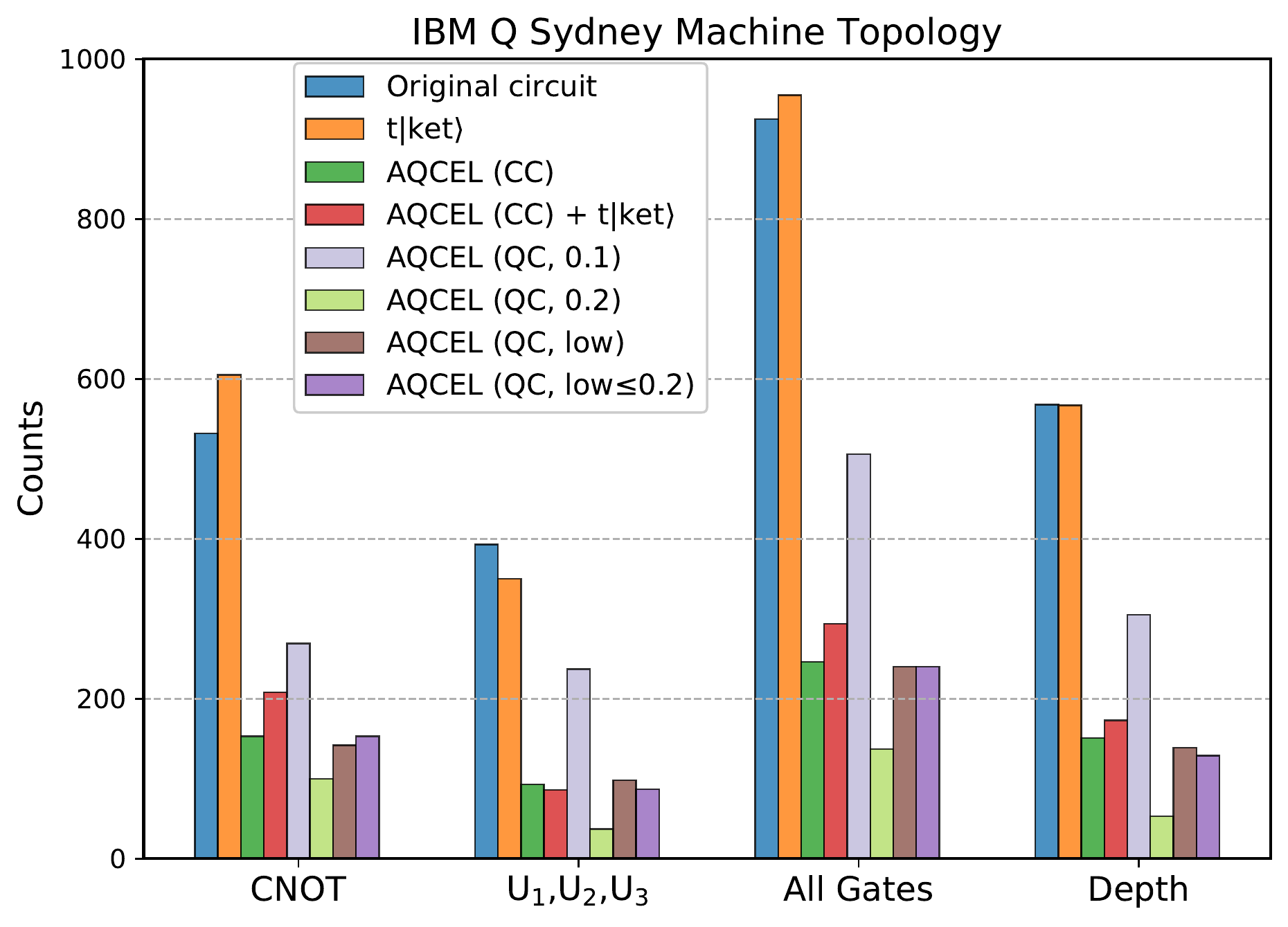}
\caption{Numbers of single-qubit ($U_{1,2,3}$) gates, CNOT gates and the sum of the two 
as well as the depth of the one-branching step QPS circuit transpiled considering \textit{ibmq\_sydney} 
topology before and after optimization under different schemes.}
\label{fig:gatecounts_1step_ps_all} 
\end{figure}

\begin{figure*}
\centering
\includegraphics[width=0.48\textwidth]{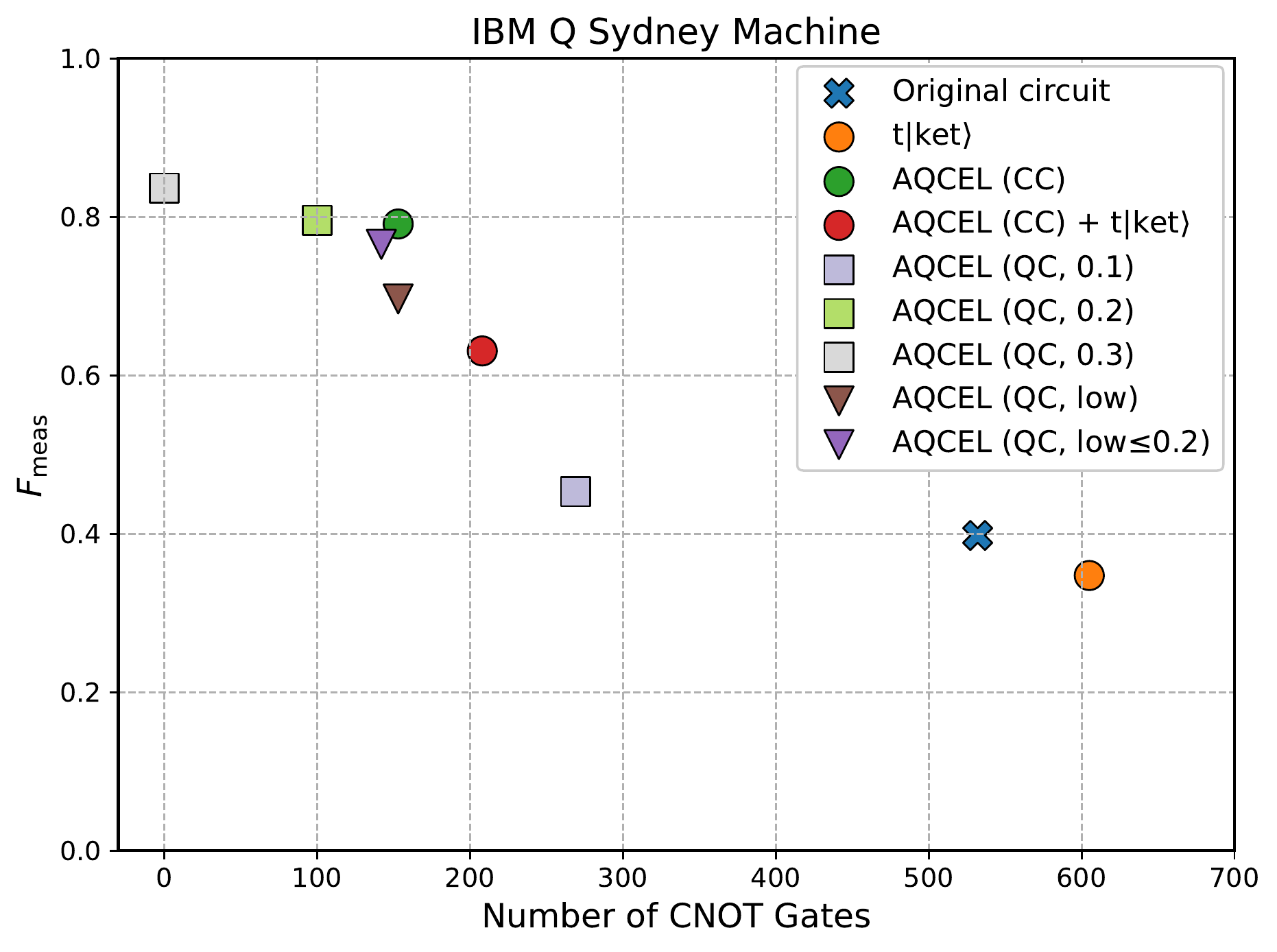}
\includegraphics[width=0.48\textwidth]{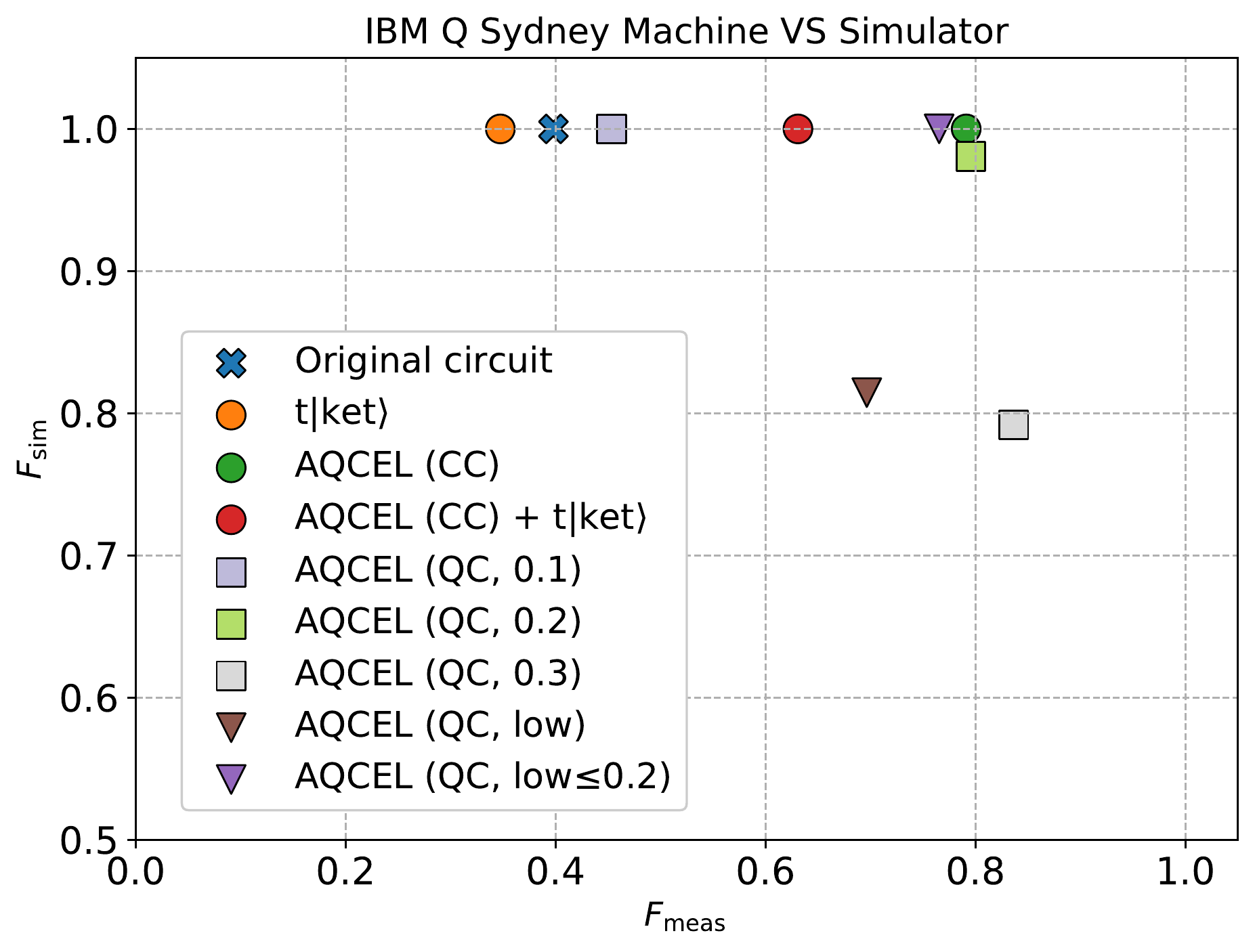}
\caption{Fidelity {\Fmeas} versus the number of CNOT gates (left) and fidelities {\Fmeas} versus {\Fsim} (right) for the one-branching step QPS circuit transpiled considering 
\textit{ibmq\_sydney} topology before and after optimization under different schemes.
These transpiled circuits are executed on \textit{ibmq\_sydney} to obtain the {\Fmeas} and a statevector simulator to obtain the \Fsim.}
\label{fig:fidelity_1step_ps_all} 
\end{figure*}

The results obtained from different approaches for finding nonzero-amplitude basis states and
different choices of cutoff thresholds are summarized in Figs.~\ref{fig:gatecounts_1step_ps_all}
and \ref{fig:fidelity_1step_ps_all} for comparison.  It is worth noting that most of
the \myopt-based optimization shown in the figure improve the {\Fmeas} value over the \tket-only
optimization.  Another interesting finding is that the determination of bitstring probabilities with
quantum measurements brings a better gate reduction than the identification of nonzero amplitudes with
classical calculation, if the cutoff threshold is set properly (0.2 for this case). A qualitative
explanation for this would be that the quantum measurements and the cutoff serve to remove qubit
controls over low-amplitude basis states, where such states contribute little to the final probability distributions. 
An exact identification of nonzero-amplitude computational basis states with
classical simulation does not lead to the removal of such qubit controls.
In addition, the determination with quantum measurements can suppress the contributions 
from spurious low-amplitude states due to the existence of hardware noise, making the {\Fmeas} value comparable to 
the one from the determination using classical calculation. 
Figure~\ref{fig:fidelity_1step_ps_all} shows that, with the proper choice of the thresholds, e.g., \sfrac of 0.2 or \slow capped at 0.2, one can make {\Fmeas} 
comparable to the case with the optimization performed using classical calculation while keeping {\Fsim} at unity. 

\section{Discussion}
\label{sec:discussion}
\subsection{Applicability of proposed heuristic optimization}
\label{diss:applicability}

The core component of the proposed heuristic circuit optimization is the identification of 
computational basis states with nonzero amplitudes and the subsequent elimination of redundant controlled operations.
Therefore, \myopt is expected to work more efficiently for quantum algorithms 
in which the quantum state has a small number of high-amplitude computational basis states. In other words, if all the computational 
basis states have non-negligible amplitudes, \myopt would not be effective.
An example of when \myopt is not effective is a quantum algorithm where an equal superposition state is first created by applying $H^{\otimes n}$ 
to the initial $\ket{0}^{\otimes n}$ state of the $n$-qubit system, 
such as Quantum Phase Estimation~\cite{nielsen00} and Grover's Algorithm.

\subsection{Possibility of further simplifications}
\label{diss:gate_reduce}

For certain quantum circuits, there is a case where there are successive multi-qubit controlled gates acting with
the same control qubits. One example is in the QPS simulation circuit (Fig.~\ref{fig:showercircuit}).
The circuit determines if an emission happens and which particle radiates or splits, depending on the total counts of particles of each type.
These steps (corresponding to the blocks with controlled unitary operations denoted by 
$U_e^{(m)}$ and $U_h$ in Fig.~\ref{fig:showercircuit}) require a lot of successive multiple controlled operations that share 
the same control qubits.
In this case, if the circuit is expanded by adding an ancilla qubit and the triggering decision of the control qubits 
is stored into the ancilla qubit, the remaining multi-qubit controlled gates can be controlled by the ancilla. 
A potential caveat is that adding ancilla qubits might introduce additional SWAP gates when implementing the circuit to hardware.
However, since this approach does not depend on the amplitudes of computational basis states of a given circuit state, 
it is complementary to the \myopt optimization scheme and will open the possibility of reducing the overall gate counts further.

Another interesting possibility is that if a circuit turns out to contain only a small number of basis states, the circuit state can 
be represented using fewer qubits than the original ones. Given that this might require a completely new computational basis, this is left for future work.

\subsection{Implication to hardware implementations of quantum circuits}
\label{diss:hardware}

The techniques introduced in the \myopt protocol, i.e., identification of most-frequently-appearing sets 
of quantum gates as RSGs and the removal of redundant qubit control operations, have 
implications to hardware implementation of quantum circuits. 

First, the RSGs would be a prioritized target for better mapping to quantum hardware. For the QPS
algorithm, the RSGs contain multi-qubit controlled gates like the Toffoli gate, as shown in
Fig.~\ref{fig:rsg_2step_ps}.  In this case, these RSGs are further decomposed into collections of
native single- and two-qubit gates.  Therefore, the depth of the transpiled circuit depends
significantly on which hardware qubits the decomposed RSG gates are mapped on to.  If the
tranpilation algorithm accounts for the frequency of the occurrence of the RSGs, an improved qubit mapping can
be created such that frequently-used controlled gates are applied on neighboring qubits with better
connectivities on the quantum hardware.

In comparison between the \myopt and \tket optimizers (e.g.,
Figs.~\ref{fig:gatecounts_2step_ps} and \ref{fig:gatecounts_1step_ps}), the \tket performance on the
gate reduction turns out to be sub-optimal for the QPS algorithm.  This is largely due to the lack
of ability in \tket to remove redundant controlled operations through the identification of
nonzero-amplitude computational basis states.  However, in certain cases, the \tket-optimized circuit
ends up with even more gates than the original circuit, as seen in
Fig.~\ref{fig:gatecounts_1step_ps} (note that the original and \tket-optimized circuits are both
{\it optimized} using the noise-adaptive mapping and gate cancellation, see
Sec.~\ref{subsec:setup}).  The \tket optimizes a circuit assuming that all the qubits are connected
to each other.  This indicates that the circuit optimized with this assumption could result in more SWAP gates
once the hardware connectivity is taken into account~\footnote{This transpilation is performed with tools in Qiskit Terra. The \tket also allows its own circuit compilation to specific hardware, but it is not examined in this study.}.
This clearly indicates that it is beneficial for removing unnecessary controlled operations as much as possible without the assumption of full qubit connectivity. Moreover, if a circuit is mainly composed of Level 3 RSGs,
as in the case of the QPS circuit used here, the hardware quality of control qubits of the RSGs will become crucial for 
the circuit simplification procedure in the \myopt protocol.

\section{Conclusion and outlook}
\label{sec:conclusion}
We have proposed a new protocol, called \myopt, for analyzing quantum circuits to identify recurring
sets of gates and remove redundant controlled operations. The heart of the redundant controlled operations removal
resides in the identification of zero- or low-amplitude computational basis states. In particular,
this procedure can be performed through measurements using a quantum computer in polynomial time,
instead of classical calculation that scales exponentially with the number of qubits. Although
removing qubit controls triggered in low-amplitude states will produce a circuit that is
functionally distinct from the original one, it is observed that this may be a desirable feature in some
cases under the existence of hardware noise.
If a quantum circuit contains recurring sets of quantum gates, those gates will be considered as candidates for further 
optimization in terms of both gate synthesis and hardware implementation. In the proposed protocol, the underlying technique to identify recurring gate sets
is demonstrated, leading to the possibility of hardware-aware optimization of such gates including dedicated microwave pulse controls.

We have explored the \myopt optimization scheme using the quantum parton shower simulation, a prototypical
quantum algorithm for high-energy physics. For this algorithm, the proposed scheme shows a
significant reduction in gate counts with respect to \tket, which is one of the industry-standard
optimization tools, while retaining the accuracy of the probability distributions of the final
state.

This feature opens the possibilities to extend this optimization scheme further in future. We have considered several 
scenarios of the thresholds applied to the measured bitstrings to take into account the gate errors. 
The measurement error is accounted for using the calibration matrix approach, and this can be improved by adapting 
the unfolding technique developed in Ref.~\cite{Bauer:2019uf} and related approaches that use fewer resources~\cite{geller_efficient_2020,song_10-qubit_2017,gong_genuine_2019,wei_verifying_2020,hamilton2020scalable} or further mitigate the errors~\cite{2010.07496}. 
A substantial contribution to the gate errors originates from CNOT gates.  
There are a variety of approaches to mitigate these errors, including the zero noise extrapolation mentioned in Sec.~\ref{sec:intro}.
The method based on the fixed identify insertion technique has been tested, showing 
that the circuit optimization improves with lower thresholds to determine the bitstring probabilities.  The random identity insertion protocol introduced in Ref.~\cite{PhysRevA.102.012426} may further reduce the gate count and thus improve the fidelity of our approach.
The threshold choice has a large impact to the accuracy of measuring the probability distributions,
as in Fig.~\ref{fig:fidelity_1step_ps_all}, therefore the precise control of the measurement and gate errors is crucial 
for this approach.

\begin{acknowledgements}
We acknowledge the use of IBM Quantum Services for this work. The views expressed are those of the authors, and do not reflect the official policy or position of IBM or the IBM Quantum team.

CWB and BN are supported by the U.S. Department of Energy, Office of Science under contract DE-AC02-05CH11231.  In particular, support comes from Quantum Information Science Enabled Discovery (QuantISED) for High Energy Physics (KA2401032).

We would like to thank Ross Duncan and Bert de Jong for useful discussions about the ZX-calculus.

\end{acknowledgements}

\appendix
\section{Algorithms of graph pattern recognition}
\label{app:pattern_recognition_alg}

The pattern recognition algorithm of recurring set of quantum gates~(RSG) is described in Algorithm 2.
This algorithm is based on depth-first search with heuristic pruning.

\begin{figure}
\centering
\includegraphics[width=0.23\textwidth]{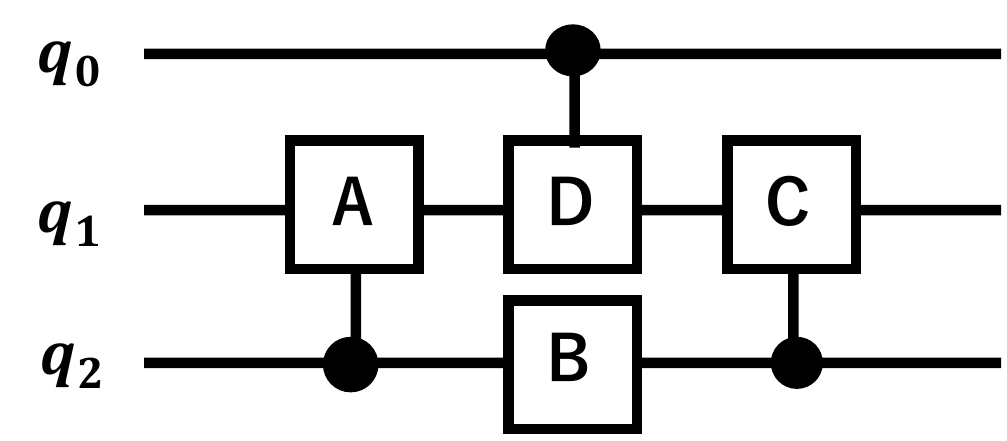}
\includegraphics[width=0.23\textwidth]{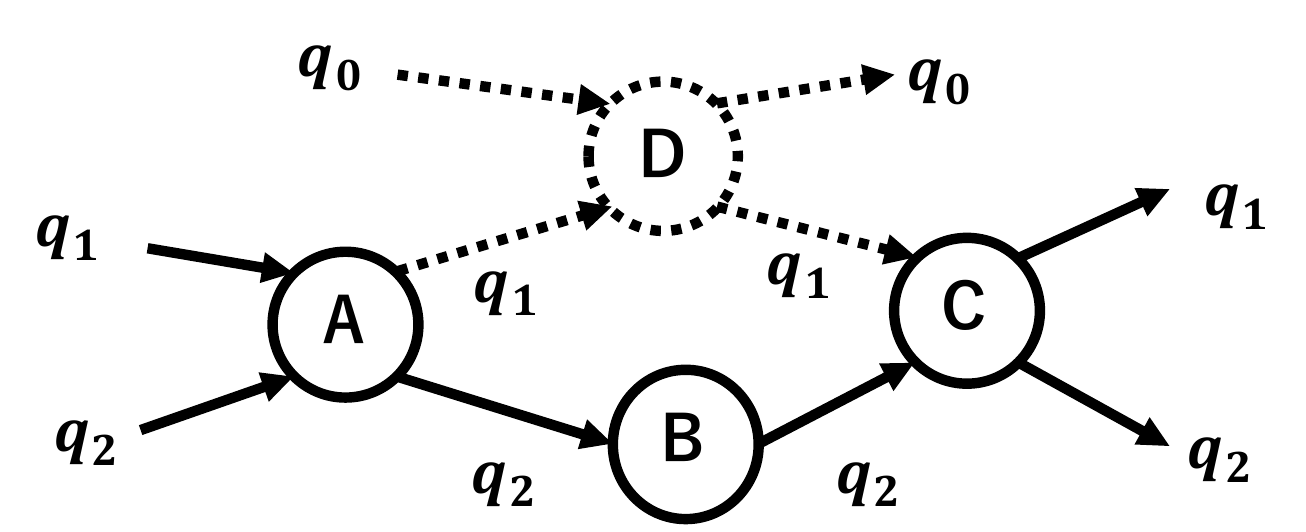}
\caption{An example of quantum circuit (left) and its subgraph~($G^{'}=\{A, B, C\}$) removed in our pattern recognition algorithm~(right). A functionality of the corresponding circuit depends on the intermediate gate~$(D)$).}
\label{fig:app:bad_DAG}
\end{figure}

First, RSG candidates are built from seeding a quantum gate~(node) by seeking possible combinations of RSGs 
that have descending connected quantum gates.
A target node used as a seed, i.e., the beginning node, is selected with postorder traversal with a memorization technique 
to avoid a repeating calculation.
The computational complexity of the algorithm is $O(N_\text{nodes}!)$~\footnote{The $i$-th node has $N_\text{nodes} - i$ RSG candidates in the worst case. Therefore, the computational complexity of the combination of the number of child-node's RSG candidates is $O(N_\text{nodes}!)$.}.
Due to a large number of combinations of recurring gates, the complexity is worse than the typical complexity of a classical computer, $O(n_\text{qubits}!)$ or $O(2^{n_\text{qubits}})$, because of $N_\text{nodes} = n_\text{gates} \geq n_\text{qubits}$ in most cases, and therefore it loses the benefit of quantum computer.
To reduce the computational complexity,
we prune the RSG candidates by requiring the length of the longest path, the minimum number and the maximum number of elements in RSG.
The requirement of the minimum number of elements rejects a trivial RSG~(e.g. $G=\{X \}$).
The computational complexity reduces to $O(N_\text{nodes}^{N_\text{thr}})$~\footnote{We take $N_\text{thr}$ RSG candidates from $N_\text{nodes}$ nodes. Therefore, the computational complexity is $\binom{N_\text{nodes}}{N_\text{thr}} \approx N_\text{nodes}^{N_\text{thr}}$.}
where $N_\text{thr}$ is a threshold value for the pruning, and the classical computer can calculate this  
in polynomial time when $N_\text{thr}$ is fixed.
However, this algorithm sometimes causes ill-defined RSGs, as shown in Fig.~\ref{fig:app:bad_DAG}.
The functionality of the quantum circuit from such an RSG depends on the intermediate gate that is not used in the RSG.
These RSGs are rejected in this algorithm by requiring that there is no node, which is both a child 
and a parent nodes but not an element of the RSG~($\exists g_i, g_j \subseteq G^{'}, \{ g_k | g_i \rightarrow g_k, g_k \rightarrow g_j  \} \nsubseteq G^{'}$).

After building the RSG candidates, they are grouped by graph isomorphism using the Weisfeiler Lehman graph hash.
The use of graph hash does not ensure that two graphs are isomorphic, but the accuracy is sufficient for our use case.
For the Level 1 matching criteria which consider only gate types, we assign the gate type as a node feature and assign nothing for an edge feature.
For the Level 2 matching criteria which consider both gate types and qubit \textit{roles}, we assign the gate type as a node feature and assign the target or control label as an edge feature.
For the Level 3 matching criteria which consider gate types, qubit \textit{roles} and \textit{indices}, we assign the gate type as a node feature and assign the absolute qubit index as an edge feature.

Finally, the top-$k$ RSGs are selected based on the frequency times the graph size.

\noindent\makebox[\linewidth]{\rule{\columnwidth}{0.8pt}}
\textbf{Algorithm 2}:  Gate set pattern recognition with DAG\vspace{-2mm}\\
\noindent\makebox[\linewidth]{\rule{\columnwidth}{0.4pt}}\vspace{-6mm}
\begin{algorithmic}
\FORALL{quantum gate~(node)~($g_i$) in the circuit~($G$)}
  \FORALL{subset~($G^{'}$) beginning with the target node~($g_i$)}
    \IF{the longest path is longer than the threshold}
      \STATE{continue}
    \ENDIF
    \IF{number of elements in subset is out of thresholds}
      \STATE{continue}
    \ENDIF
    \IF{$\exists g_i, g_j \subseteq G^{'}, \{ g_k | g_i \rightarrow g_k, g_k \rightarrow g_j  \} \nsubseteq G^{'} $}
      \STATE{continue}
    \ENDIF
    \STATE{$G^{'}$ is a RSG candidate.}
  \ENDFOR
\ENDFOR

\STATE{Make a set of RSGs~($S(h) = \{ G^{'} | \mathrm{hash}(G^{'}) = h \}$)}
\STATE{Select top-$k$ sets of RSGs~(S(h)) ordering by the frequency~$|S(h)|$ times RSG size~($|G^{'}|$)}
\label{app:alg:RSGalg}
\end{algorithmic}
\vspace{-2mm}
\noindent\makebox[\linewidth]{\rule{\columnwidth}{0.4pt}}

\section{General conditions to eliminate qubit controls}
\label{app:qubit_control}

Given a multi-qubit controlled gate $C^{m}[U]$ and a system in the ``undetermined'' state
$\ket{\psi}$ following the classification in Section~\ref{subsubsec:basic_idea}, we can derive the
condition for removal of a part of the controls to be allowed in the following way.

Let $x$ be the
number of controls to be removed. Without loss of generality, the decomposition of $\ket{\psi}$ can
be rewritten as
\begin{equation}
  \ket{\psi} = \sum_{i, l, k} \tilde{c}_{i, l, k} \ket{i}_{\text{ctl}'} \otimes \ket{l}_{\text{free}} \otimes \ket{k},
\end{equation}
where $\ket{\cdot}_{\text{ctl}'}$ and $\ket{\cdot}_{\text{free}}$ are the states of the $m-x$
remaining control qubits and the $x$ qubits from which the controls are removed. From Eq.~\eqref{eq:state_decomposition},
\begin{equation}\label{eq:basis_breakdown}
  \ket{i}_{\text{ctl}'} \otimes \ket{l}_{\text{free}} = \ket{2^{x} i + l}_{\text{ctl}},
\end{equation}
and therefore
\begin{equation}\label{eq:coeff_identity}
  \tilde{c}_{i, l, k} = c_{2^{x}i + l, k}.
\end{equation}

Applying the original controlled gate to $\ket{\psi}$ yields
\begin{multline}\label{eq:controlled_original}
  C^{m}[U] \ket{\psi} = \sum_{j=0}^{2^{m} - 2} \sum_{k} c_{j, k} \ket{j} \ket{k} + \\
  \sum_{k} c_{2^{m} - 1, k} \ket{2^{m} - 1} U \ket{k},
\end{multline}
where ket subscripts and the tensor product symbols are omitted for simplicity.
In contrast, the new gate with fewer controls gives
\begin{multline}\label{eq:controlled_reduced}
  C^{m-x}[U] \ket{\psi} = \sum_{i=0}^{2^{m-x} - 2} \sum_{l,k} \tilde{c}_{i, l, k} \ket{i} \ket{l} \ket{k} + \\
  \sum_{l, k} \tilde{c}_{2^{m-x} - 1, l, k} \ket{2^{m-x} - 1} \ket{l} U \ket{k}.
\end{multline}

For the removal of $x$ qubit controls to be allowed, the right hand sides of
Eqs.~\eqref{eq:controlled_original} and \eqref{eq:controlled_reduced} must be identical. This requires
\begin{multline}\label{eq:state_identity}
  \sum_{l=0}^{2^{x} - 2} \sum_{k} \tilde{c}_{2^{m-x} - 1, l, k} \ket{2^{m - x} - 1} \ket{l} U \ket{k} = \\
  \sum_{l=0}^{2^{x} - 2} \sum_{k} c_{2^{m} - 2^{x} + l, k} \ket{2^{m} - 2^{x} + l} \ket{k}.
\end{multline}
Denoting
\begin{equation}
  U \ket{k} = \sum_{k'} u_{kk'} \ket{k'}
\end{equation}
and recalling Eq.~\eqref{eq:basis_breakdown}, Eq.~\eqref{eq:coeff_identity}, Eq.~\eqref{eq:state_identity} implies (replacing $k' \leftrightarrow k$ on the left hand side)
\begin{multline}
  \sum_{l=0}^{2^{x} - 2} \sum_{k,k'} \tilde{c}_{2^{m-x} - 1, l, k'} u_{k'k} \ket{2^{m - x} - 1} \ket{l} \ket{k} = \\
  \sum_{l=0}^{2^{x} - 2} \sum_{k} \tilde{c}_{2^{m-x} - 1, l, k} \ket{2^{m - x} - 1} \ket{l} \ket{k}.
\end{multline}
Then, we have
\begin{multline}\label{eq:the_condition}
  \sum_{k'} \tilde{c}_{2^{m - x} - 1, l, k'} u_{k'k} = \tilde{c}_{2^{m - x} - 1, l, k} \\
  \quad \forall l \in \{0, 1, ..., 2^x - 2\}, k.
\end{multline}
Eq.~\eqref{eq:the_condition} holds if the row vector $\{\tilde{c}_{2^{m - x} - 1, l, k}\}_{k}$
is an eigenvector of the matrix $u$ with eigenvalue 1 under right multiplication for $0 < l < 2^x - 2$, or if
$\tilde{c}_{2^{m - x} - 1, l, k} = 0$ for $0 < l < 2^x - 2$ and all $k$.

Since the cost of exactly computing the complex amplitudes of the quantum state is high, in \myopt
we only consider this second condition:
\begin{equation}\label{eq:aqcel_condition}
  \tilde{c}_{2^{m - x} - 1, l, k} = 0 \quad \forall l \in \{0, 1, ..., 2^x - 2\}, k.
\end{equation}
When using quantum measurements to estimate the bitstring probabilities at the control qubits, this requirement
corresponds to observing no bitstring with 1 in all control qubits, except when $l = 2^x - 1$.
In other words, there should be no bitstring by which $C^{m}[U]$ is not triggered but $C^{m-x}[U]$ is.

\section{Computational resources for the proposed optimization scheme}
\label{app:comp_resource}

The computational cost needed to perform the proposed optimization scheme is evaluated here.
We consider a quantum circuit that contains $n$ qubits and $N$ multi-qubit controlled gates, each 
acting on $m$ control qubits and one target qubit.

The elimination of adjacent gate pairs proceeds, for each gate, by checking a pair-wise matching 
to the next gate until the end of the gate sequence. Since the gate can act on at most $n$ qubits, 
the computational cost is ${\cal O}(nN)$.

The next step in the optimization scheme is the identification of computational basis states.
If we use the classical calculation for simply tracking all the computational basis states
whose amplitudes may be nonzero at each point of the circuit without the calculation of the amplitudes,
it requires the computation of ${\cal O}(N2^n)$ states, so the resource requirement grows exponentially with $n$.
This method requires less computational resource than a statevector simulation 
but it neglects certain rare cases where exact combinations of amplitudes lead to the elimination of redundant controlled operations.
If we measure the control qubits at each controlled gate $M$ times using a quantum computer,
the total number of gate operations and measurements is given by
$M\{m+(1+m)+(2+m)+\cdots+(N-1+m)\} = \frac12MN(N-1)+mMN$.
Therefore, the computational cost grows as ${\cal O}(MN^2+mMN)$, i.e., polynomially with $n$.

We next consider removing redundant qubit controls from a controlled gate with $m$ control qubits.
Using a quantum computer that measures the $m$ control qubits $M$ times,
the measured number of bitstrings is $M$ if $M<2^m$, otherwise $2^m$.
For the classical calculation, the number of basis states is $2^m$. 
Imagine that we choose an arbitrary combination among $2^m$ possible combinations of new qubit controls on the same controlled gate. 
If we want to know 
whether the chosen combination can act as the correct qubit control, we need to check, for a given measurement 
done previously with a quantum computer, if all measured bitstrings satisfy Eq.~\eqref{eq:aqcel_condition}. 
This requires ${\cal O}(m2^m)$ checks for one bitstring.
Since this has to be checked for all the measurements, the cost is ${\cal O}(Mm2^m)$ if $M<2^m$, 
otherwise ${\cal O}(m4^m)$ for one chosen combination. Therefore, the overall computational cost for the determination of redundant qubit controls 
is ${\cal O}(Mm4^mN)$ or ${\cal O}(m8^mN)$ for $N$ multi-qubit controlled gates, each 
having $2^m$ combinations of new qubit controls. The classical calculation requires ${\cal O}(m8^mN)$ as well.

It is known that an arbitrary multi-qubit controlled-$U$ gate with $m$ control qubits can be decomposed into 
${\cal O}(m)$ Toffoli and two-qubit controlled-$U$ gates~\cite{PhysRevA.52.3457}. 
Therefore, if a controlled gate in the circuit is 
decomposed in this way, then above computational cost for the redundant qubit controls would become 
${\cal O}(mN)$. With this decomposition, the total number of 
gate operations and measurement increases due to ${\cal O}(m)$ times more controlled gates. 
However, the computational cost for the identification of computational basis states becomes only $\frac12mMN(mN-1)+2mMN$, 
so it still behaves polynomially as ${\cal O}(m^2MN^2)$ when quantum computer is used. 
For the classical calculation, the cost becomes ${\cal O}(mN2^n)$.

The final step of the optimization scheme is the elimination of unused qubits. This is performed by simply
checking qubits that all the gates in the circuit act on, corresponding to a computational cost of ${\cal O}(nN)$.

Given that a controlled gate has at most $n-1$ control qubits, 
the total computational cost for the entire optimization sequence is ${\cal O}(n^2MN^2)$ or ${\cal O}(nN2^n)$, depending on
whether the computational basis state measurement is performed using a quantum computer or a classical calculation.

\bibliographystyle{plainnat}
\bibliography{cirq_opt}

\end{document}